\documentclass[%
 aip,
 jmp,%
 amsmath,amssymb,
 floatfix,
 reprint,%
]{revtex4-1}

\usepackage{graphicx}
\usepackage{dcolumn}
\usepackage{bm}
\newcommand{\mathLaplace}{\Delta}

\newcommand{\tmop}[1]{\ensuremath{\operatorname{#1}}}

\newcommand{\tmtextbf}[1]{\text{{\bfseries{#1}}}}

\newcommand{\codestar}[1]{\texttt{#1}}

\begin{document}


\title{Effects of resonant magnetic perturbations on neutral beam heating in a tokamak}

\author{Youjun Hu}
\affiliation{Institute of Plasma Physics, Chinese Academy of Sciences, Hefei 230031, China}%

\author{Yingfeng Xu}%
 \email{xuyingfeng@dhu.edu.cn}
\affiliation{College of Science, Donghua University, Shanghai 201620, China}%

\author{Baolong Hao}%
 \affiliation{Advanced Energy Research Center, Shenzhen University, Shenzhen 518060, China}%

\author{Guoqiang Li}
\author{Kaiyang He}
\author{Youwen Sun}
\affiliation{Institute of Plasma Physics, Chinese Academy of Sciences, Hefei 230031, China}%

\author{Li Li}
\affiliation{College of Science, Donghua University, Shanghai 201620, China}%

\author{Jinfang Wang}
\author{Juan Huang}
\author{Lei Ye}
\author{Xiaotao Xiao}
\affiliation{Institute of Plasma Physics, Chinese Academy of Sciences, Hefei 230031, China}%
\author{Feng Wang}
\affiliation{School of Physics, Dalian University of Technology, Dalian 116024, China}%

\author{Chengkang Pan}
\author{Yongjian Xu}
\affiliation{Institute of Plasma Physics, Chinese Academy of Sciences, Hefei 230031, China}%

\date{\today}

\begin{abstract}
  Effects of resonant magnetic perturbations (RMPs) on tangential neutral beam
  heating in the EAST tokamak are studied numerically. RMPs with linear
  resistive magnetohydrodynamics response are used in the modeling. A variety
  of representing configurations of RMP coil currents are examined and their
  effects on the NBI heating efficiency are compared, in order to find a
  parameter window where deleterious effects of RMPs on NBI heating efficiency
  are minimized. It is found that the internal redistribution of fast ions by
  RMPs induces local accumulation of fast ions, resulting in higher local fast
  ion pressure than the case without RMPs. It is also found that the toroidal
  phasing of the RMP with respect to the fast ion source has slight effects on
  the steady-state radial profile of fast ions. The dependence of fast ion
  loss fraction on the RMP up-down phase difference shows similar behavior as
  the dependence of the radial width of chaotic magnetic field on the phase
  difference. A statistical method of identifying resonances between RMPs and
  lost fast ions is proposed and the results indicate that some resonances
  between RMPs and lost passing particles may be of non-integer fractional
  order, rather than the usual integer order.
\end{abstract}

\maketitle

\section{Introduction}
Neutral beam injection (NBI) is widely adopted in tokamaks for heating plasma
and driving flow{\cite{Scoville2019,Hu2015,Schneider_2011,ASUNTA201533,Heidbrink_2009}}.
Modeling the neutral beam heating, which involves neutral particle ionization
and fast ion collisional transport, is a mature field where simulation results
usually agree reasonably with experiments{\cite{pankin2004}}. One of the
uncertainties in the modeling is how electromagnetic perturbations, if
present, affect the heating and fast ion transport process. These
electromagnetic perturbations can be various intrinsic modes such as
Alfv{\'e}n eigenmodes, and can also be externally imposed perturbations such
as resonant magnetic perturbations (RMPs).

This work numerically studies the influence of RMPs on the Deuterium neutral
beam heating in the EAST tokamak{\cite{Wan_2017,Wan_2019}}. EAST is equipped
with RMP coils which are designed for the control of edge localized modes
(ELMs){\cite{Sun_2016,sun2016prl}}. This coils turn out to also have effects
on the transport of neutral beam fast
ions{\cite{Xu_2020,yxu2019,He_2019,he2020a,he2020b}}. The effects of RMPs on
fast ion transport and loss have been extensively investigated on AUG
{\cite{Garcia_Munoz_2013,Garcia_Munoz_2013b,Sanchis_2018}},
KSTAR{\cite{kim2018}}, DIII-D{\cite{Van_Zeeland_2013,Van_Zeeland_2015}}, and
MAST{\cite{McClements_2015}} tokamaks.

In this work, we consider RMPs of $n = 1$ with linear resistive
magnetohydrodynamics (MHD) plasma response, where $n$ is the toroidal mode
number of the RMP coil current. The configurations with $n = 0, 2, 3, 4$ were
also examined (not shown in this paper). The results indicate that $n = 0$ RMP
has negligible effects on the confinement of fast ions, as is expected, and
the deleterious effects of configurations with $n = 2, 3, 4$ on the
confinement of fast ions are generally smaller than that of the $n = 1$ RMP.
Therefore we will focus on the $n = 1$ configuration in this paper.

We consider one of the four neutral beams on EAST that is injected in the
co-current direction, with the tangent radius being $1.26 m$ and the full
(kinetic) energy being $51 \tmop{keV}$. The number ratio between neutrals of
the full energy, half energy and $1 / 3$ energy is chosen to be $80 : 14 : 6$.
NBI power after neutralizing is assumed to be 1MW. Further details on the
neutral beam modeling are provided in Appendix \ref{21-4-14-p1} and
\ref{21-4-14-p2}. In this paper, fast ions are defined as those ions that are
born from the neutral beam ionization and have not yet been slowed down to the
energy $2 T_{i 0}$, where $T_{i 0}$ is the temperature of background thermal
ions at the magnetic axis.

We found that the internal redistribution of fast ions by RMPs can induce
local accumulation of fast ions, resulting in larger local fast ion pressure
than the case without RMPs. It is also found that the toroidal phasing of the
RMP with respect to the fast ion source has slight effects on the steady-state
radial profile of fast ions. The dependence of fast ion loss fraction on the
RMP up-down phase difference $\Delta \Phi$ is found to show similar behavior
as the dependence of the radial width of chaotic magnetic field on $\Delta
\Phi$.

A simple statistical method for identifying resonances between RMPs and lost
fast ions is proposed, in which resonances are identified by examining peaks
in the graph of $p$ versus $\Delta f$, where $p = n \omega_{\phi} /
\omega_{\theta}$ is called resonance order with $\omega_{\phi}$ and
$\omega_{\theta}$ being the toroidal angular frequency and poloidal angular
frequency of lost fast ions, respectively, and $\Delta f$ is the difference of
the lost fast ion distribution function between the case with RMP and that
without RMP. By taking the difference, we exclude all the fast ions that are
lost due to the first-orbit prompt loss and pure collision loss. The results
indicate that some resonances may be of non-integer fractional order, rather
than the usual integer order.

In the process of studying the effects of RMP on NBI heating, we developed a
new Monte-Carlo code which models continuous neutral beam injection,
ionization, and the resulting fast ions transport under the influence of RMPs.
This code (referred to as {\codestar{TGCO}}) is similar to the established NBI
modeling code {\codestar{NUBEAM}}{\cite{pankin2004}} and many other test
particle orbit-following codes such as {\codestar{OFMC}}{\cite{Tani1981}},
{\codestar{ASCOT}}{\cite{HIRVIJOKI2014}},
{\codestar{ORBIT}}{\cite{White_2010}},
{\codestar{SPIRAL}}{\cite{Kramer_2013}},
{\codestar{VENUS-LEVIS}}{\cite{PFEFFERLE20143127}},
{\codestar{GYCAVA}}{\cite{yxu2019}}, {\codestar{SOFT}}{\cite{he2020b}}.
{\codestar{TGCO}} has been used as a NBI module in Ref.
{\cite{Xu_2020,he2020a,xyXU_2020}}. The guiding-center drift model, edge loss
model, and fast ion collision model adopted in this work are well known, and
the details of these models are provided in the Appendix. The finite Larmor
radius (FLR) effect is taken into account when pushing particle guiding center
drift, depositing markers to compute density and pressure, and checking
whether particles are lost to the boundary. The simulation includes the
X-points of equilibrium magnetic field and uses the first wall as the particle
loss boundary. For simplicity, we assume a pure Deuterium plasma with no
impurities.

The remainder of this paper is organized as follows. Section \ref{19-12-26-5}
describes the equilibrium configuration, bulk plasma profiles, and neutral
beam ionization profiles. Section \ref{21-5-30-a1} introduces the RMP coils on
EAST and the resulting 3D magnetic perturbations with/without plasma response.
Section \ref{21-6-2a1} discusses effects of RMPs on the steady state radial
profiles of neutral beam heating and fast ion pressure. Section \ref{21-6-2a4}
examines the dependence of volume integrated heating power, fast ion stored
energy, and fast ion loss fraction on the up-down phase difference of RMP coil
currents, where radial width of chaotic magnetic region is computed and is
found to be closely related to the fast ion loss fraction. Section
\ref{21-6-2a5} use a statistical method to identify the resonance between RMPs
and fast ions. A brief summary and some discussions are given in Sec.
\ref{19-12-26-8}.

\section{Equilibrium and fast ion source}\label{19-12-26-5}

\begin{figure}[htp]
  \includegraphics[scale=0.5]{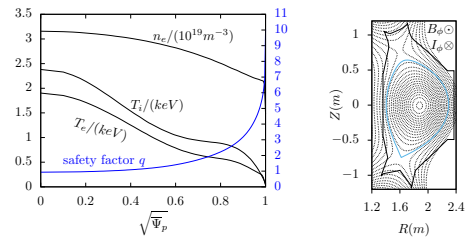}
  \caption{\label{17-2-22-p1m}Left panel: profiles of electron number density
  $n_e$ , electron temperature $T_e$, ion temperature $T_i$, and safety factor
  $q$ of EAST discharge \#52340@3.4s. The radial coordinate $\rho =
  \sqrt{\overline{\Psi}_p}$ is the square root of the normalized poloidal
  magnetic flux, with $\overline{\Psi}_p = (\Psi - \Psi_0) / (\Psi_b -
  \Psi_0)$, where $\Psi \equiv A_{\phi} R$ is the poloidal flux function,
  $\Psi_0$ and $\Psi_b$ are the values of $\Psi$ at the magnetic axis and
  last-closed-flux-surface, respectively. Right panel: magnetic configuration.
  This is a low single null configuration with $B_{\tmop{axis}} = 2.2 T$, $I_p
  = 404 \tmop{kA}$, $q_{\tmop{axis}} = 0.95$, and $q_{95} = 5.05$.}
\end{figure}

EAST is a superconducting tokamak with a major radius $R_0 = 1.85 m$, minor
radius $a \approx 0.45 m$, typical on-axis magnetic field strength
$B_{\tmop{axis}} \approx 2.2 T$ and plasma current $I_p \approx 0.5
\tmop{MA}${\cite{Wan_2017,Wan_2019}}. The magnetic configuration and plasma
profiles used in this work (Fig. \ref{17-2-22-p1m}) were reconstructed by EFIT
code from EAST tokamak discharge \#52340@3.4s with constrains from experiment
diagnostics.

The neutral beam ionization is modeled by the Monte-Carlo method and the
details are provided in Appendix \ref{21-5-30-a3}. Figure \ref{17-4-13-p1}
plots the two-dimensional distribution of ionized particles in the poloidal
plane (averaged over the toroidal direction) and in the toroidal plane
(averaged over the vertical direction). The results indicate most fast ions
are born on the low-field-side.

\begin{figure}[htp]
  \includegraphics[scale=0.75]{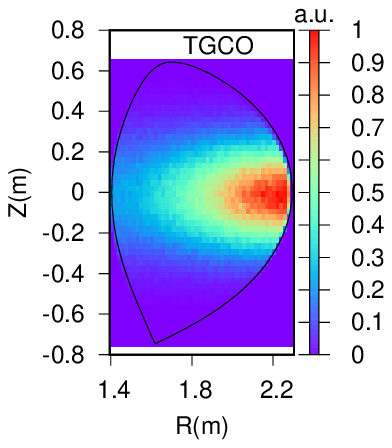}\includegraphics[scale=0.7]{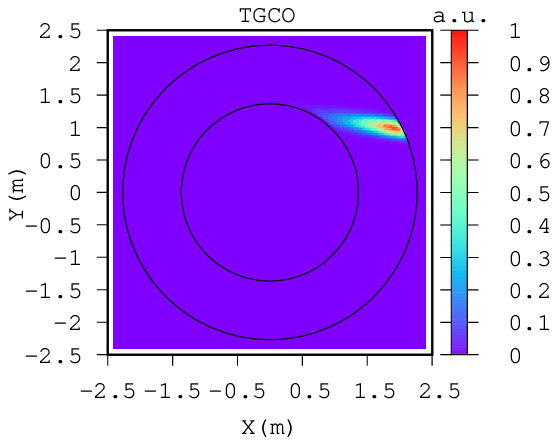}
  \caption{\label{17-4-13-p1}Neutral beam ionization profiles in the poloidal
  plane (left) and toroidal plane (right) computed by {\codestar{TGCO}}. This
  is for the neutral beam injection with $R_{\tan} = 1.26 m$ in EAST discharge
  \#52340@3.4s. The loss fraction of neutral particles to the inner first wall
  (shine-through loss) is 14\% in this case.}
\end{figure}

\begin{figure}[htp]
  \includegraphics[scale=0.6]{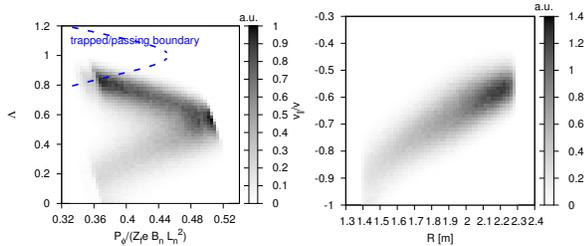}
  \caption{\label{20-9-14-e1}Distribution of ionized particles in $(\Lambda,
  P_{\phi})$ plane (left) and $(\lambda, R)$ plane (right), where $\Lambda =
  \mu B_{\tmop{axis}} / \varepsilon$, $\lambda = v_{\parallel} / v$, $P_{\phi}
  = m_f g v_{\parallel} / B_0 + Z_f e \Psi$ is the canonical toroidal angular
  momentum, $\mu$ is the magnetic moment, $v_{\parallel}$ is the parallel (to
  the magnetic field) velocity, $B_0$ is the equilibrium magnetic field, $m_f$
  and $Z_f e$ are the mass and charge of fast ions, respectively, $g =
  B_{\phi} R$, $B_n = 1 T$, $L_n = 1 m$. The trapped/passing boundary in
  $(\Lambda, P_{\phi})$ plane is indicated (the region to the left side of the
  curve is the trapped region).}
\end{figure}

Figure \ref{20-9-14-e1} plots the distribution of ionized particles in
$(\Lambda, P_{\phi})$ plane and $(\lambda, R)$ plane, where $\Lambda = \mu
B_{\tmop{axis}} / \varepsilon$, $\lambda = v_{\parallel} / v$, $\varepsilon$
is ion kinetic energy, $P_{\phi}$ is the canonical toroidal angular momentum.
The trapped/passing boundary in $(\Lambda, P_{\phi})$ plane is plotted, which
indicates that most ions born in this case are in the passing region, with a
small number of ions being near the trapped/passing boundary.

\section{Three dimensional magnetic perturbations}\label{21-5-30-a1}

\begin{figure}[htp]
  \includegraphics[scale=0.6]{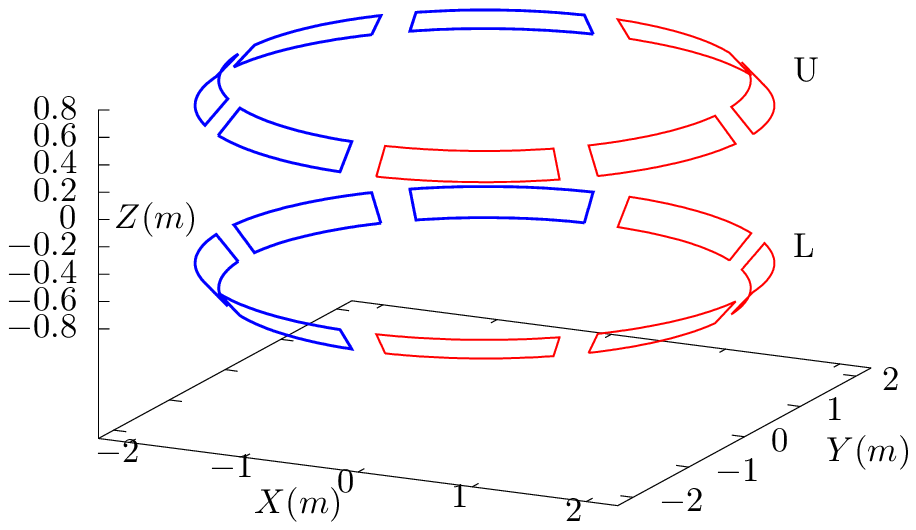}\includegraphics[scale=0.6]{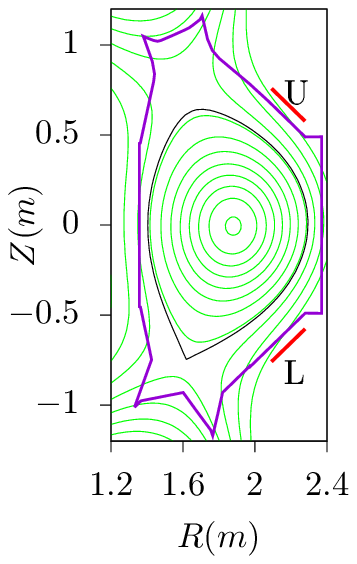}
  \caption{\label{19-9-24-1}RMP coils on EAST tokamak in 3D view (left) and
  poloidal view (right). Each coil has 4 turns with maximum current $2.5
  \tmop{kA}$ per turn. The maximum frequency of alternating current (AC)
  operation is $1 \tmop{kHz}$ (usually operating at less than $10 \tmop{Hz}$).
  This paper considers only direct current (DC) operation, i.e. static RMP. In
  the left panel, red color denotes current circulating in the clockwise
  direction and blue denotes current circulating in the anti-clockwise
  direction, viewed from the outside of the torus. The magnitude of currents
  are set to be the same. The setup shown here is called $n = 1$ and
  $\mathLaplace \Phi = 0$ configuration, where $\Delta \Phi$ is the up-down
  phase difference.}
\end{figure}

EAST has 16 RMP coils consisting of two arrays of 8 coils with up-down
symmetry located on the low field side and uniformly distributed along the
toroidal direction. Figure \ref{19-9-24-1} shows the setup of the RMP coils on
EAST tokamak{\cite{Sun_2016}}.

The vacuum magnetic perturbations produced by the RMP coils are numerically
calculated by using the Biot-Savart law. The magnetic perturbations including
the the effect of plasma response are computed, using the linear resistive MHD
model, by MARS-F code {\cite{liu2000,liu2010}} as a boundary value problem,
taking into account of a given toroidal flow. The RMP coil current, located in
the vacuum region outside the plasma, is directly modeled in the code as a
source term {\cite{liu2010}}.

\begin{figure}[htp]
  \includegraphics[scale=0.55]{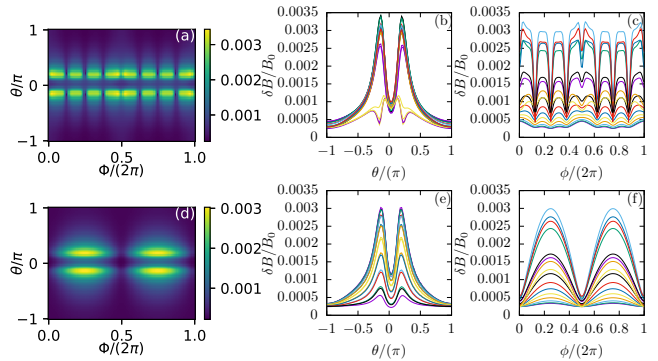}
  \caption{\label{21-11-15-p1}(a) and (d): Two-dimensional distribution of
  $\delta B / B_0$ on the $\rho = 0.96$ magnetic surface. (b) and (e):
  Poloidal dependence of $\delta B / B_0$ at a series of toroidal locations at
  $\rho = 0.96$. (c) ad (f): Toroidal dependence of $\delta B / B_0$ at a
  series of poloidal locations at $\rho = 0.96$. The upper panels are for the
  full vacuum field and the lower panels are for filtered vacuum field
  (keeping only $n = \pm 1$ Fourier harmonics). $\delta B = | \delta
  \mathbf{B} |$ is the magnetic perturbation magnitude and $B_0$ is the local
  value of the equilibrium magnetic field. $\phi$ is the toroidal angle in the
  $(R, \phi, Z)$ cylindrical coordinates. The poloidal angle $\theta$ is
  chosen to be of the PEST type{\cite{cheng1987}}, i.e., magnetic field lines
  are straight in $(\theta, \phi)$ plane (the positive direction of $\theta$
  is chosen to be counterclockwise when viewed along $\nabla \phi$ direction;
  $\theta \in [- \pi, \pi]$ with $\theta = - \pi$ being in the high field side
  midplane.) The two peaks of $\delta B / B_0$ in the poloidal direction
  appears near the poloidal locations of the RMP coils. The RMP up-down
  phasing $\Delta \Phi = 0$ in this case.}
\end{figure}

In this work, we use the $n = 1$ RMP with a current amplitude of $10
\tmop{kAt}$ unless otherwise specified. To perceive the magnitude of the
resulting magnetic perturbation (without plasma response) relative to the
equilibrium field, Figures \ref{21-11-15-p1}(a,d) show the two-dimensional
distribution of $\delta B / B_0$ on the $\rho = 0.96$ magnetic surface. Figs.
\ref{21-11-15-p1}(b,e) show the poloidal dependence of $\delta B / B_0$ at a
series of toroidal locations at $\rho = 0.96$ and Figs. \ref{21-11-15-p1}(c,f)
show the toroidal dependence of $\delta B / B_0$ at a series of poloidal
locations at $\rho = 0.96$. The results indicate that the maximal value of
$\delta B$ at $\rho = 0.96$ is about 0.3\% of the equilibrium field $B_0$. The
corresponding minimal value of $\delta B$ is small (0.03\% of $B_0$) but is
not exact zero.

The magnetic perturbation component that is perpendicular to the 2D
equilibrium magnetic surface (i.e., the radial component) is of most interest
to magnetic confinement since it creates magnetic islands at resonant surfaces
and thus changes the topology of magnetic field. Define the normalized radial
component of the magnetic perturbation by
\begin{equation}
  \delta B_N = \frac{\delta \mathbf{B} \cdot \nabla \rho}{\mathbf{B}_0 \cdot
  \nabla \phi} .
\end{equation}
Fourier expansion of $\delta B_N$ is written as
\begin{eqnarray}
  \delta B_N (\rho, \theta, \phi) & = & \sum_{n = - \infty}^{\infty} \delta
  B_N^{(n)} (\rho, \theta) \exp [- i n \phi] \\
  & = & \sum_{n = - \infty}^{\infty} \sum_{m = - \infty}^{\infty} \delta
  B_N^{(m n)} (\rho) \exp [i (m \theta - n \phi)], 
\end{eqnarray}
where $m$ and $n$ are the poloidal and toroidal mode numbers, respectively,
$\delta B_N^{(n)}$ and $\delta B_N^{(m n)}$ are the Fourier expansion
coefficients.

\begin{figure}[htp]
  \includegraphics[scale=0.7]{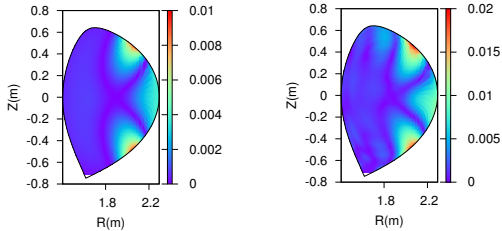}
  \caption{\label{20-12-29-p1}Amplitude of $n = 1$ harmonic of $\delta B_N$,
  i.e., $| \delta B_N^{(1)} |$, in the poloidal plane. The left panel is for
  the vacuum RMP and the right panel is for the response RMP.}
\end{figure}

Figure \ref{20-12-29-p1} compares the amplitude of $n = 1$ harmonic of $\delta
B_N$ in the poloidal plane between the vacuum RMP and the response RMP. The
results indicates that the amplitude of perturbation are enhanced by the
plasma response at some poloidal locations, which might be related to the kink
response{\cite{boozer2001prl}}. The results also show that there is also some
minor perturbation appearing near the top region in the case of response RMP
and there is twist effect on the distribution of $| \delta B_N^{(1)} |$ by the
plasma response.

\begin{figure}[htp]
  \includegraphics[scale=0.7]{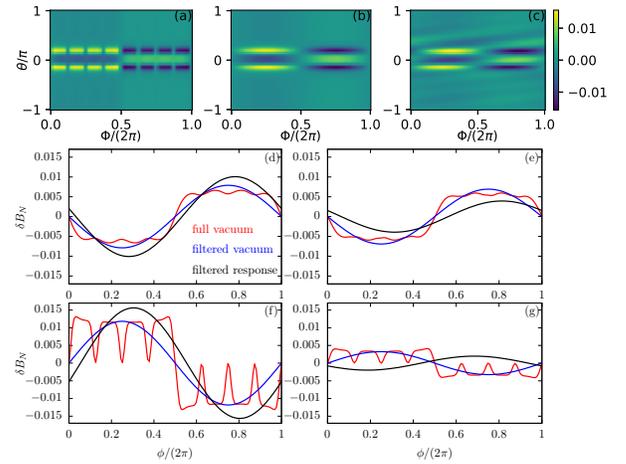}
  \caption{\label{20-12-14-a1}Distribution of magnetic perturbation $\delta
  B_N$ over $(\phi, \theta)$ on the $\rho = 0.96$ magnetic surface. (a)-(c)
  are for the two-dimensional distributions. (d)-(g) are for the
  one-dimensional distributions along $\phi$ at fixed values of poloidal angle
  $\theta$. Specifically, (d) is for $\theta = 0$, (e) is for $\theta = 0.09
  \pi$, (f) is for $\theta = 0.18 \pi$, and (g) is for $\theta = 0.29 \pi$.
  Three kinds of perturbations are shown: magnetic perturbation produced by
  the coils (full vacuum), the filtered vacuum field, and the filtered
  perturbation when the plasma response is considered (filtered response). (a)
  is for the full vacuum field, (b) is for the filtered vacuum field, and (c)
  is for the filtered field with plasma response.}
\end{figure}

Figure \ref{20-12-14-a1} shows the distribution of magnetic perturbation
$\delta B_N$ over $(\phi, \theta)$ on the $\rho = 0.96$ magnetic surface.
Three kinds of perturbations are shown: the full vacuum RMP, the filtered
vacuum RMP (keeping only the $n = \pm 1$ Fourier components) and the filtered
response RMP. One-dimensional distributions along $\phi$ at fixed values of
poloidal angle $\theta$ are also show in Fig. \ref{20-12-14-a1} to compare the
amplitude and phase along the toroidal direction between the three kinds of
magnetic perturbations. The results indicate that the amplitude of
perturbation are enhanced by the plasma response at at some poloidal locations
(Fig \ref{20-12-14-a1}d, f) but are slightly reduced at others (Fig.
\ref{20-12-14-a1}e). The results also indicate that the response RMP has a
toroidal phase shift relative to the vacuum one, which ranges from minor
difference (in Fig. \ref{20-12-14-a1}d) to major difference (anti-phasing in
Fig. \ref{20-12-14-a1}g).

\begin{figure}[htp]
  \includegraphics[scale=0.7]{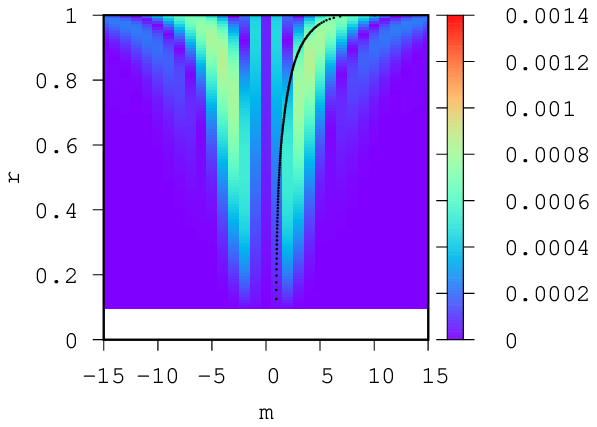}\includegraphics[scale=0.7]{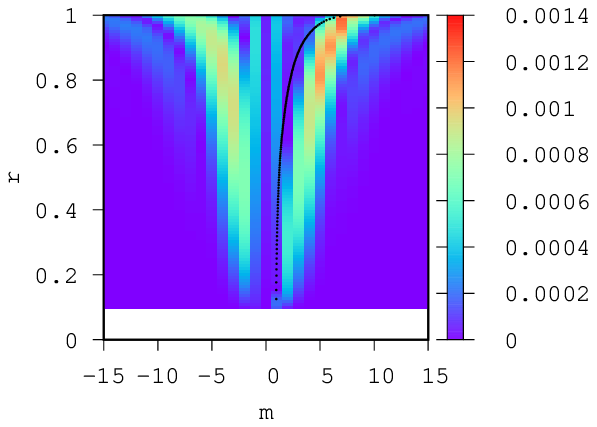}
  \caption{\label{20-12-28-a1}Amplitude of the poloidal Fourier spectrum of
  the $n = 1$ harmonic of $\delta B_N$ for different radial locations, i.e.,
  $| \delta B_N^{(m n)} (\rho) |$ with $n = 1$. The left panel is for the
  vacuum field and the right panel is for the field with plasma response. RMP
  coil configuration of $\Delta \Phi = 0$ is used. The black curves are $m = n
  q$, which indicate the resonant locations.}
\end{figure}

To clearly distinguish the resonant components from the non-resonant ones, \
Figure \ref{20-12-28-a1} gives the heatmap of Fourier components of $\delta
B_N$ in the $(m, \rho)$ plane. Figure \ref{20-12-28-a1} compares the poloidal
Fourier spectrum of the $n = 1$ harmonic of $\delta B_N$ between the vacuum
RMP and response RMP. The main difference between vacuum and response RMPs is
that plasma response reduces the perturbation amplitude at resonant locations
(shielding effects) and enhances the amplitude of non-resonant components, as
is shown in Fig. \ref{20-12-28-a1}, where the resonant locations $m = n q$ are
indicated.

\begin{figure}[htp]
  \includegraphics{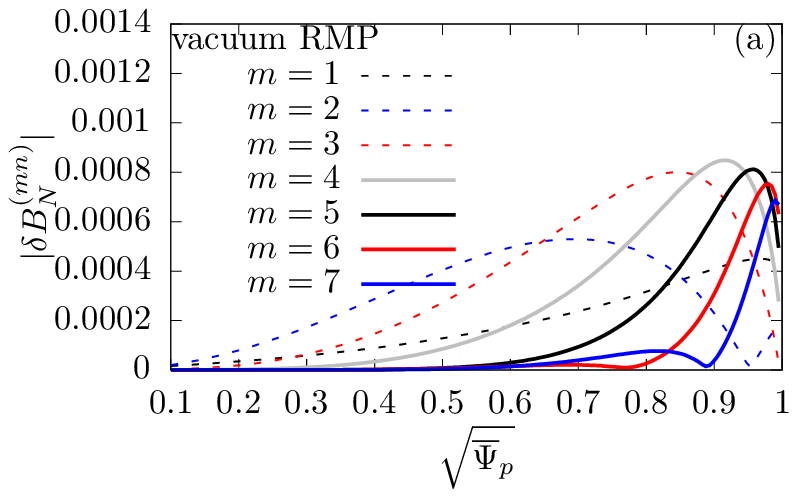}
  \includegraphics{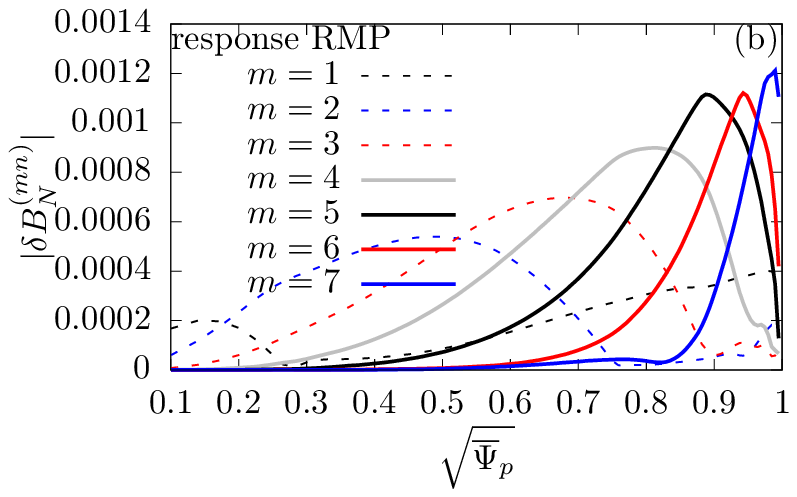}
  \caption{\label{21-11-4-a1}Radial profile of amplitude of various poloidal
  Fourier harmonics of $n = 1$ RMPs of phasing $\Delta \Phi = 0$ for the
  vacuum case (a) and plasma response case (b). Only $m > 0$ harmonics are
  shown here. The differences of the $m \leqslant 0$ harmonics between the
  vacuum RMP and response RMP are minor (which can be roughly recognized from
  Fig. \ref{20-12-28-a1}) and are thus not shown here.}
\end{figure}

To more clearly show the effects of plasma response, Figure \ref{21-11-4-a1}
compares the radial profiles of amplitude of poloidal harmonics between vacuum
RMP and response RMP. The results indicate that the peaks of the harmonics are
shifted inward to the core by the plasma response and the peak values near the
edge are enhanced. The results also show that plasma response reduces the
magnitude of harmonics near their respective resonant locations (e.g., $m = 2$
at $\rho = 0.76$ and $m = 3$ at $\rho = 0.89$).

Using $\delta B_N^{(m n)}$ defined above, the width of magnetic island
generated by a magnetic perturbation at a rational surface $q = m / n \equiv
q_s$ is given by{\cite{Sun_2015}}
\begin{equation}
  \label{21-4-28-1} w_{m n} = 4 \sqrt{\left| \frac{\rho \delta B_N^{(m n)}}{n
  S} \right|_{q = q_s}},
\end{equation}
where $S \equiv \rho q' / q$ with $q' = d q / d \rho$ being the magnetic
shear. Formula (\ref{21-4-28-1}) will be used later in this paper to calculate
the Chirikov parameter to determine whether adjacent islands overlap. The
Chirikov parameter is defined by
\begin{equation}
  \sigma = \frac{w_{m_1 n} + w_{m_2 n}}{2 | \rho_2 - \rho_1 |},
\end{equation}
where $\rho_1$ and $\rho_2$ are the radial coordinates of two adjacent
rational surfaces, $w_{m_1 n}$ and $w_{m_2 n}$ are the width of magnetic
islands at the two radial locations. Then the two islands overlap if $\sigma >
1$, which further indicates that magnetic stochasticity appears.

In the remainder of this paper, RMPs refer to the RMPs with plasma response
unless otherwise specified.

\section{Modification of radial profile of ion heating power density by
RMPs}\label{21-6-2a1}

\

With continuous beam injection into a time-independent background plasma, a
steady state of fast ions can be reached on the time scale of slowing-down
time ($\sim 100 \tmop{ms}$ for typical parameters of EAST tokamaks). A steady
state is reached when there is a balance between the source (beam injection
and ionization) and sink (thermalization and edge loss). In this section, we
examine the steady state of the radial profiles of NBI heating power density
and fast ion pressure.

\begin{figure}[htp]
  \includegraphics[scale=0.75]{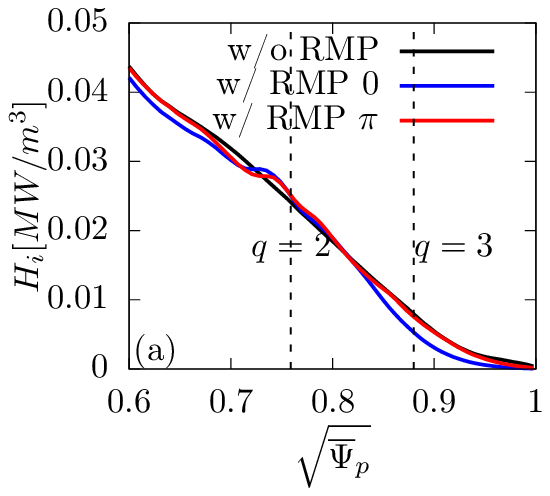}\includegraphics[scale=0.75]{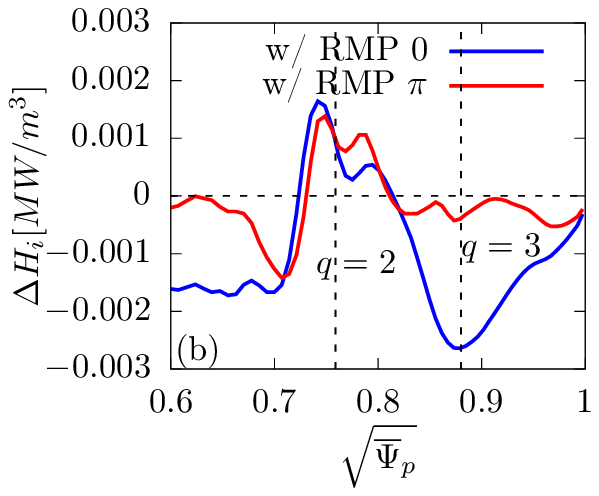}
  \caption{\label{21-4-12-p1}(a) Radial profiles of NBI heating power density
  delivered to bulk ions. (b) Difference of ion heating power density between
  the case with RMP and that without RMP. The locations of $q = 2$ and $q = 3$
  rational surfaces are indicated. The FLR effect is included in the
  simulations.}
\end{figure}

Fig. \ref{21-4-12-p1}a compares the radial profiles of the NBI heating power
density delivered to bulk ions between the case without RMP and those with
RMPs of $\Delta \Phi = 0, \pi$. To clearly show the difference, Fig.
\ref{21-4-12-p1}b plots the power density difference $\Delta H_i$, where
$\Delta H_i \equiv H_i^{(\tmop{rmp})} - H_i^{(\tmop{normp})}$ is the power
density difference between the case with RMP and that without RMP. The results
indicate that the ion heating power density is significantly reduced near the
edge (around the radial location $q = 3$) by the RMPs of $\Delta \Phi = 0$
while very marginal reduction is induced by the the RMP of $\Delta \Phi =
\pi$. The results also show that the heating power density is slightly
increased around $q = 2$ by the RMPs.

\begin{figure}[htp]
  \includegraphics[scale=0.75]{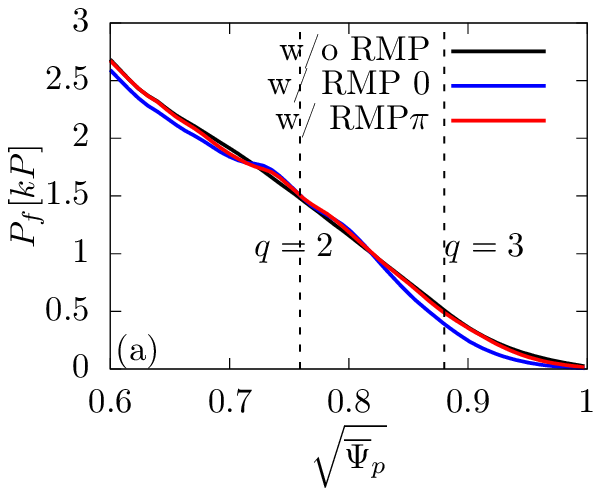}\includegraphics[scale=0.72]{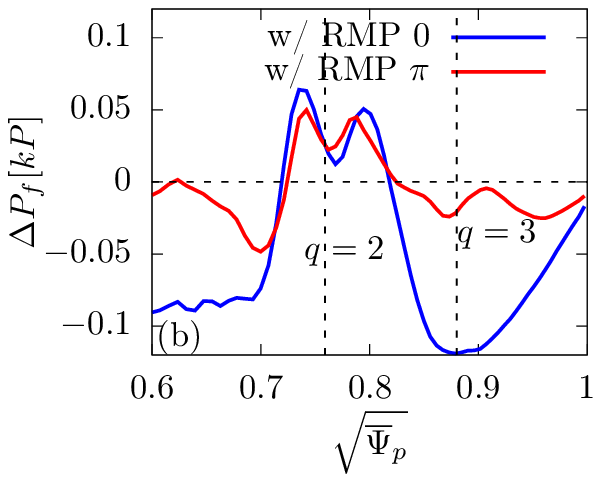}
  \caption{\label{21-5-17-a2}(a) Radial profiles of fast ion pressure. (b)
  Difference of fast ion pressure between the case with RMP and that without
  RMP.}
\end{figure}

The radial profile of NBI heating power density is determined by the radial
profile of fast ion pressure (see Appendix \ref{21-5-17-a1} for the formula of
computing the power density). Figure \ref{21-5-17-a2}a plots the radial
profile of fast ion pressure for the three cases: without RMP, with RMP of
$\Delta \Phi = 0$, and with RMP of $\Delta \Phi = \pi$. To clearly show the
difference, Fig. \ref{21-5-17-a2}b plots the pressure difference $\Delta P_f$,
where $\Delta P_f \equiv P_f^{(\tmop{rmp})} - P_f^{(\tmop{normp})}$ is the
power density difference between the case with RMP and that without RMP. The
results indicate that the influence of RMPs on fast ion pressure is similar to
that of RMPs on the ion heating power, i.e., the pressure is significantly
reduced near the edge (around the radial location $q = 3$) by the RMPs of
$\Delta \Phi = 0$ while very marginal reduction is induced by the the RMP of
$\Delta \Phi = \pi$ and the pressure is slightly increased around $q = 2$ by
the RMPs.

\begin{figure}[htp]
  \includegraphics[scale=0.65]{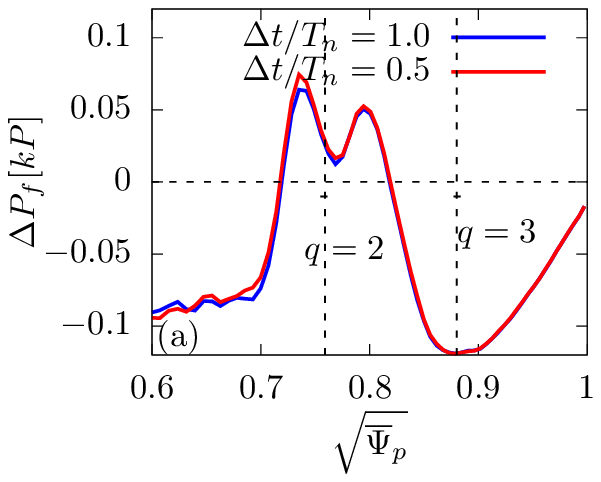}\includegraphics[scale=0.65]{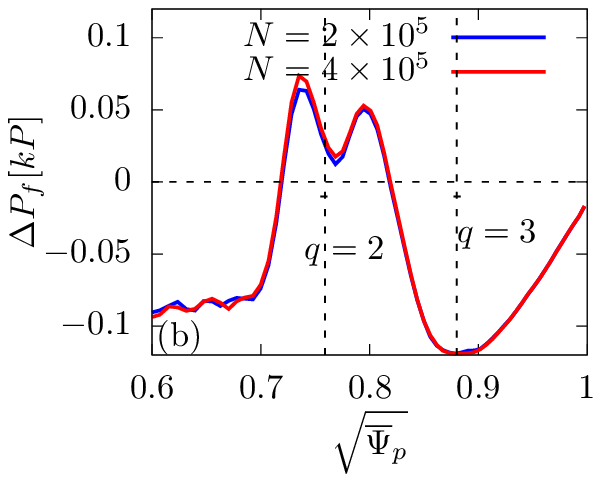}
  \includegraphics[scale=0.65]{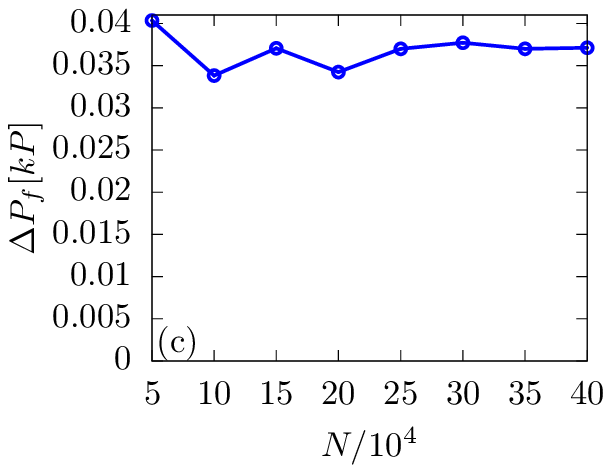}\includegraphics[scale=0.65]{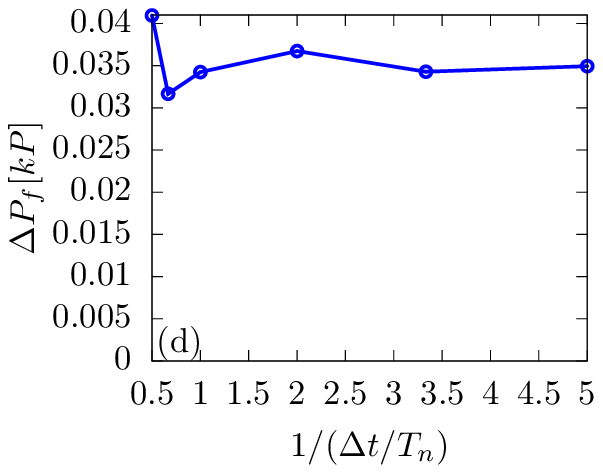}
  \caption{\label{21-5-17-a3}Comparison of radial profiles of fast ion
  pressure perturbation for different values of time step sizes (a) and marker
  number (b). Panel (c) plots the $\Delta P_f$ at $\sqrt{\overline{\Psi}_p} =
  0.755$ as a function of number of markers used in the simulation. Panel (c)
  plots the $\Delta P_f$ as a function of the inverse of the time step size $1
  / (\Delta t / T_n)$, where $T_n = 2 \pi / (B_n Z_f e / m_f)$ with $B_n = 1
  T$. RMP of $\Delta \Phi = 0$ is used in the simulations. In (a) and (d),
  marker number $N = 2 \times 10^5$. In (b) and (c), time step size $\Delta t
  / T_n = 1.0$.}
\end{figure}

As is shown in Fig. \ref{21-5-17-a2}, the pressure increasing around $q = 2$
surface induced by RMPs is quite small. It is reasonable to doubt whether the
increasing is physical or due to numerical errors. As a partial verification
of this issue, we perform convergence studies over some numerical parameters
used in the simulations such as the maker numbers and time step sizes. Figure
\ref{21-5-17-a3} shows that the radial profiles of $\Delta P_f$ agree with
each other for two different values of the time step size and marker number.
Also shown in Fig. \ref{21-5-17-a3} is the dependence of values of $\Delta
P_f$ at a specified radial location ($\rho = 0.755$) over the time step size
and marker number, which shows reasonable convergence.

It is not surprising to observe the local increasing of fast ion pressure when
a RMP is imposed. RMPs generally have pump out effects on fast ions. However
the pump out effects can be non-uniform along the radial direction, which
implies that some radial locations may accumulate fast ions, resulting higher
fast ion pressure than the no RMP case.

\begin{figure}[htp]
  \includegraphics[scale=0.8]{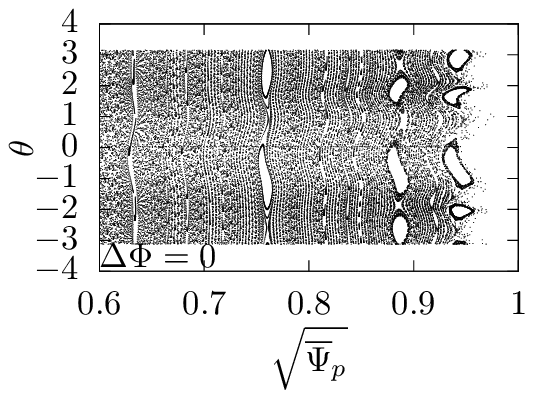}\includegraphics[scale=0.8]{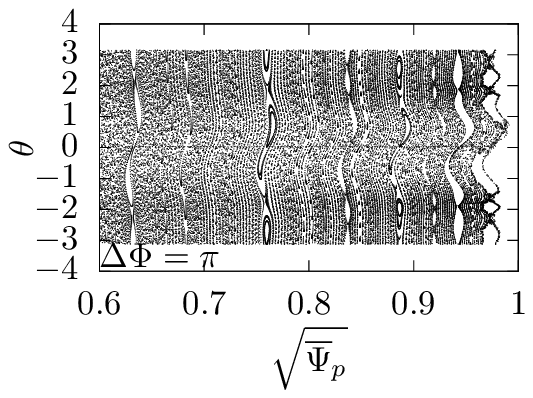}
  \caption{\label{21-4-28-p1}Comparison of magnetic field line Poincare plots
  between RMP of $\Delta \Phi = 0$ and that of $\Delta \Phi = \pi$.}
\end{figure}

\begin{figure}[htp]
  \includegraphics[scale=0.9]{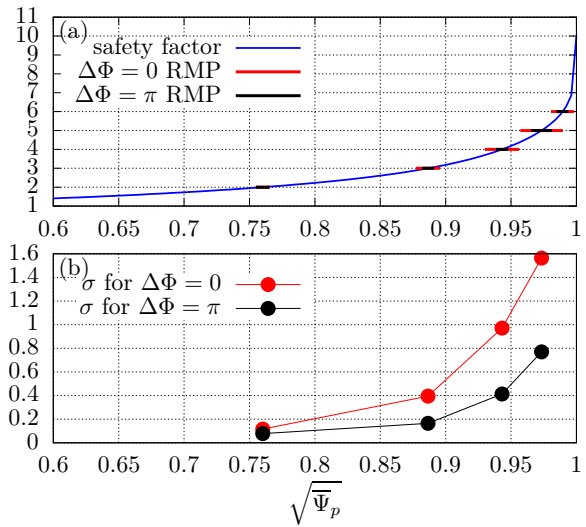}
  \caption{\label{21-4-28-p2}Magnetic island width (upper panel) and Chirikov
  parameter values (low panel) for RMPs of $\Delta \Phi = 0$ and $\Delta \Phi
  = \pi$. The safety factor profile is also shown to indicate the radial
  locations of rational surfaces. The island width is calculated by using
  formula (\ref{21-4-28-1}).}
\end{figure}

The results in Fig. \ref{21-5-17-a2}b indicate that the radial fluctuation of
fast ion pressure relative to the no RMP case depends on the RMP up-down phase
difference $\mathLaplace \Phi$: radial fluctuation for $\mathLaplace \Phi = 0$
is larger than the case of $\mathLaplace \Phi = \pi$. To understand this, we
compare the resonant and non-resonant components for the two phasings. The
amplitude of resonant components is related to the width of magnetic islands
[Eq. (\ref{21-4-28-1})], which can be inferred from Poincare sections of
magnetic field lines. Figure (\ref{21-4-28-p1}) shows the Poincare plots of
magnetic field line for RMPs of $\Delta \Phi = 0$ and $\Delta \Phi = \pi$. The
results indicate that magnetic islands for RMP of $\Delta \Phi = 0$ are wider
than those of $\Delta \Phi = \pi$. The magnetic island width can also be
directly calculated by using Eq. (\ref{21-4-28-1}). The results are plotted in
Fig. \ref{21-4-28-p2}a, which confirm the conclusion draw from the Poincare
plots.


The values of Chirikov parameter $\sigma$ at the rational surfaces $q = 2, 3,
4, 5$ are plotted in Fig. \ref{21-4-28-p2}b, which indicates $\sigma > 1$ at
$q = 5$ for $\Delta \Phi = 0$ RMP while $\sigma$ is still less than one at $q
= 5$ for $\Delta \Phi = \pi$ RMP. This implies that the radial width of
stochastic magnetic field is larger for $\Delta \Phi = 0$ RMP than that of
$\Delta \Phi = \pi$ RMP.

\begin{figure}[htp]
  \includegraphics[scale=0.9]{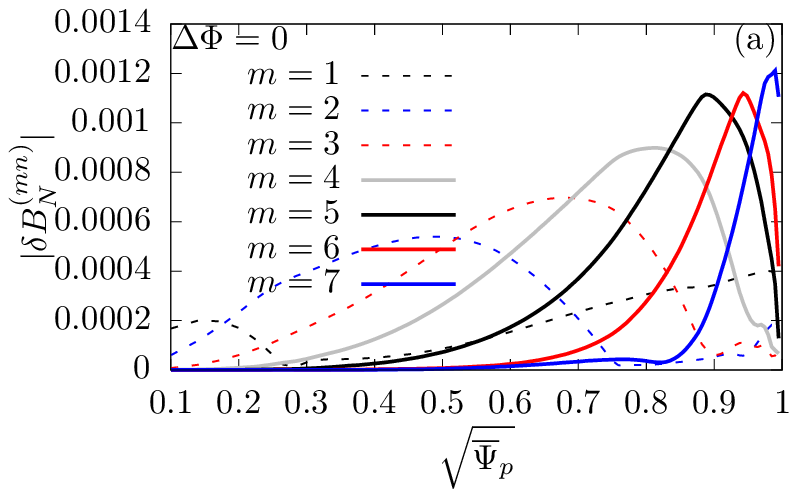}
  \includegraphics[scale=0.9]{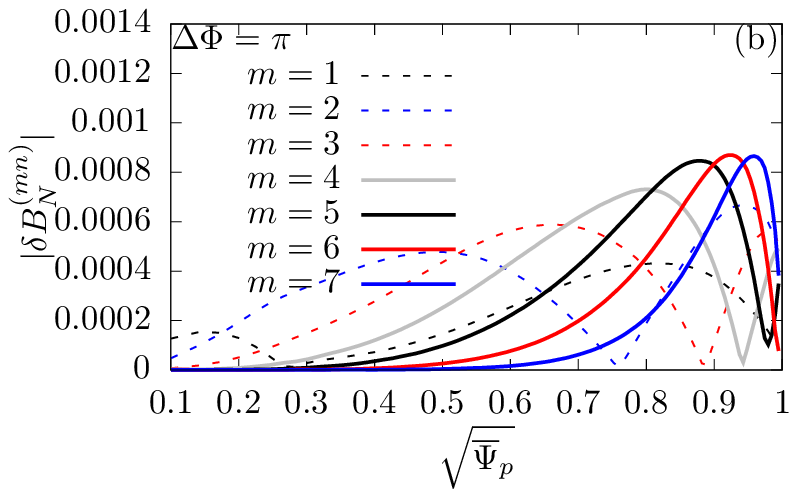}
  \caption{\label{21-5-26-p3}Radial profiles of poloidal Fourier harmonics of
  $n = 1$ RMPs of $\Delta \Phi = 0$ (a) and $\Delta \Phi = \pi$ (b).}
\end{figure}

The width of magnetic islands characterizes the amplitude of resonant
components. To characterize the non-resonant components, Fig. \ref{21-5-26-p3}
plots the radial profiles of amplitude of various poloidal Fourier harmonics.
The radial peaks of the harmonics can be used as a measure of magnitude of
non-resonant harmonics. The results clearly show that amplitude of the
non-resonant components (except for $m = 1$) for $\Delta \Phi = 0$ are larger
than those of $\Delta \Phi = \pi$.

The above results indicate that both resonant components and non-resonant
components of RMP of $\Delta \Phi = 0$ are larger than those of $\Delta \Phi =
\pi$. Both the resonant components and non-resonant components can contribute
to the redistribution of fast ions. Therefore the larger fluctuation of fast
ion pressure for the case of $\Delta \Phi = 0$ is probably due to the larger
resonant and non-resonant components for this coil current configuration.

\begin{figure}[htp]
  \includegraphics{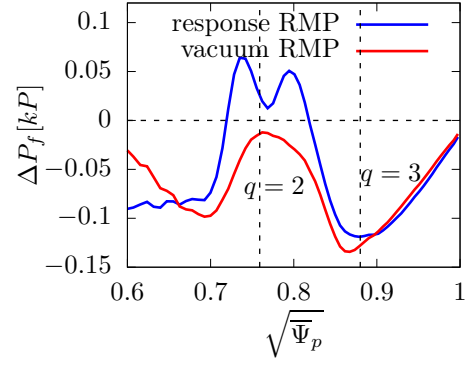}
  \caption{\label{21-11-8-e1}Difference of steady-state fast ion pressure
  between the case with RMP and that without RMP. The blue curve is for the
  RMP with plasma response while the red curve is for the vacuum RMP. Both the
  RMPs are of up-down phasing $\Delta \Phi = 0$.}
\end{figure}

Plasma response usually helps heal magnetic islands, i.e., reduces the
resonant components, as is shown in Figs. \ref{20-12-28-a1} and \
\ref{21-11-4-a1}, and thus is believed to be beneficial to fast ion
confinement{\cite{he2020b}}. Figure \ref{21-11-8-e1} compares the steady-state
fast ion profiles (subtracted by the profile when no rmp is imposed) between
the vacuum RMP and response RMP. The results shows that the fast ion pressure
around $q = 2$ for the response RMP is higher than that for the vacuum RMP,
which suggests that plasma response is beneficial to fast ion confinement
around the $q = 2$ surface. However, the results also show that the fast ion
pressure within $\sqrt{\overline{\Psi}_p} \approx 0.65$ for the response RMP
case is lower than the vacuum RMP case. This may be due to the enhancement of
$m = 1$ and $m = 2$ harmonics at their non-resonant locations by the plasma
response, as is shown by Fig. \ref{21-11-4-a1}.

\begin{figure}[htp]
  \includegraphics[scale=0.65]{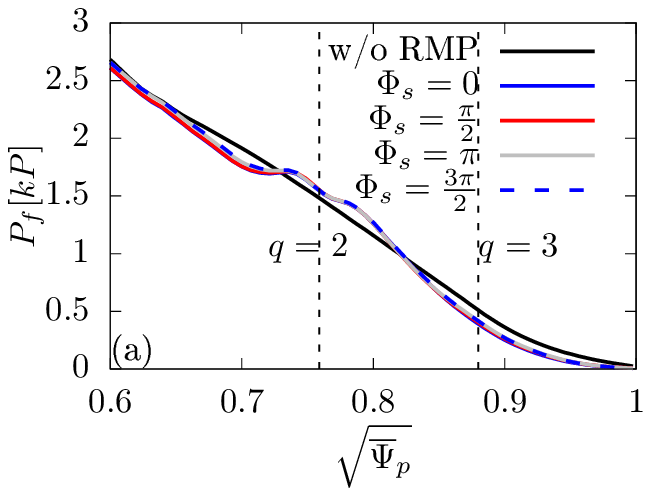}\includegraphics[scale=0.65]{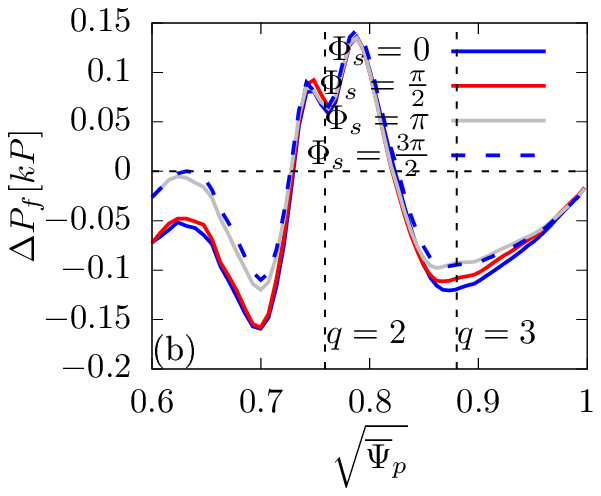}
  \caption{\label{21-5-17-a3m}(a) Radial profiles of fast ion pressure under
  RMPs of various toroidal phasings relative to the fast ion source, $\Phi_s$.
  Also shown in (a) is the radial profile when RMPs are absent. (b) The
  difference between the case with RMP and that without RMP for various values
  of $\Phi_s$. The up-down phase difference $\Delta \Phi$ is fixed at $\pi /
  2$. \ The results indicate that the toroidal phasing has slight effect on
  the steady-state radial profile of fast ions.}
\end{figure}

Since fast ion source from NBI is localized in a narrow toroidal span, it is
natural to ask whether the toroidal phasing of RMPs relative to the fast ion
source can have effects on the steady-state radial profile of fast ions. To
answer this question, we compute the steady-state radial profile of fast ion
pressure under various toroidal phasings while fixing the up-down phase
difference $\Delta \Phi$. The results are plotted in Fig. \ref{21-5-17-a3m}
for $\Delta \Phi = \pi / 2$, which indicate that the toroidal phasing has
slight effects on the steady-state radial profile. One interesting observation
is that the profiles around $q = 2$ are almost identical among the cases of
different toroidal phasings.

The effect of toroidal phasing on radial profile being slight may be valid
only for passing fast ions resulted from the tangential injection considered
in this paper. In Ref. {\cite{Sanchis_2021}}, the toroidal phasing is found to
have significant effects on the fast ion confinement, where both passing and
trapped particles are significant in the fast ion source.

\section{Up-down phase dependence of volume integrated
quantities}\label{21-6-2a4}

\begin{figure}[htp]
  \includegraphics[scale=0.5]{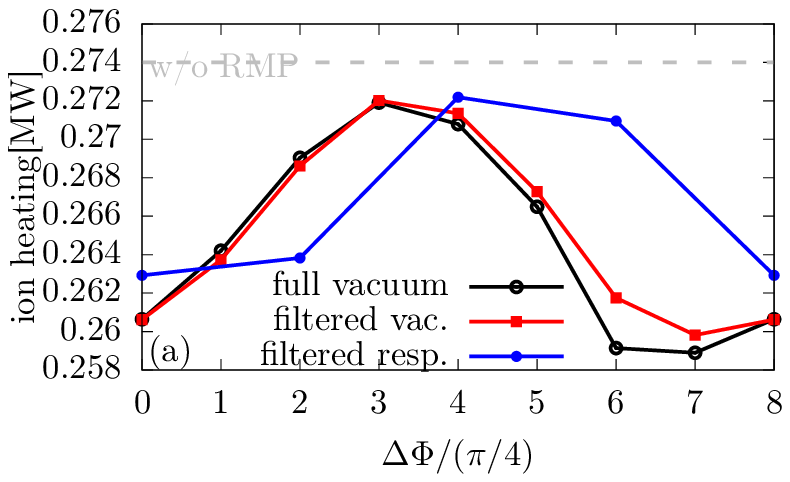}\includegraphics[scale=0.5]{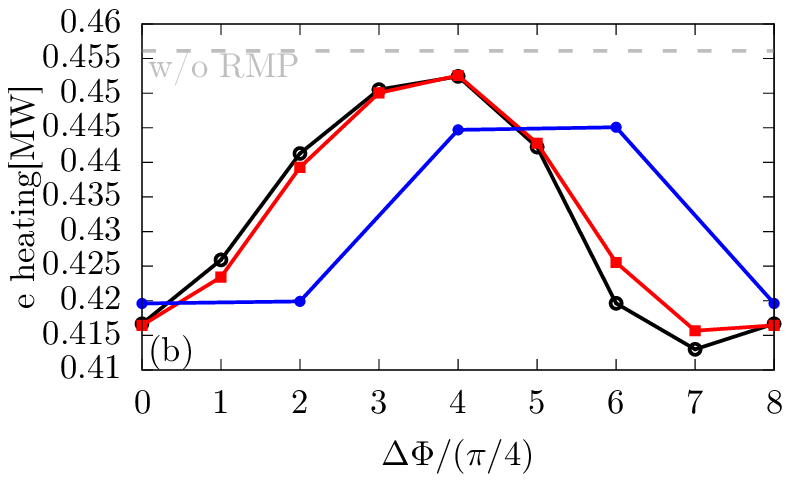}
  \includegraphics[scale=0.5]{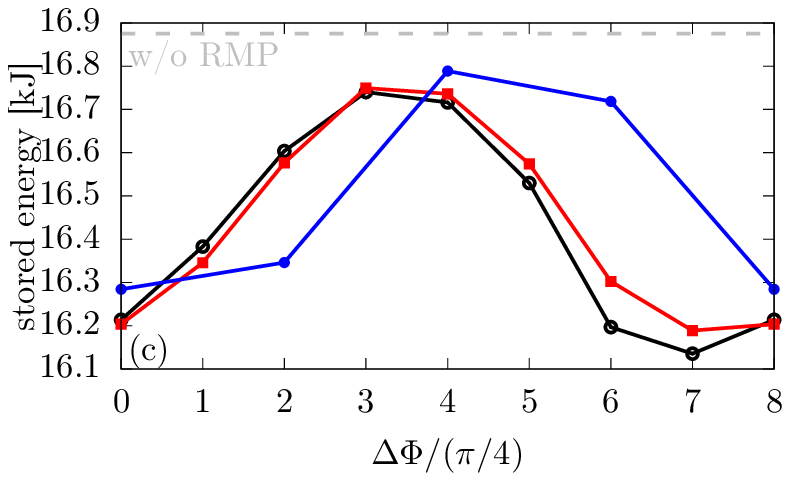}\includegraphics[scale=0.5]{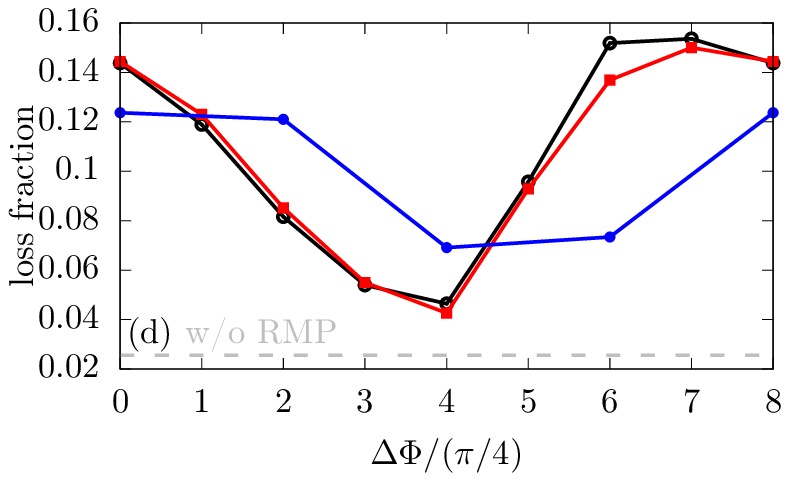}
  \caption{\label{21-5-26-a1}The up-down phasing dependence of (a) ion heating
  power, (b) electron heating power, (c) fast ion stored energy, and (d) fast
  ion loss fraction. The corresponding values in the no RMP case are shown for
  comparison. The results for three kinds of magnetic perturbations are shown,
  namely full vacuum magnetic perturbation generated by the coils (without
  filtering), filtered vacuum field (keeping only the $n = \pm 1$ Fourier
  harmonic), and the filter response RMP.}
\end{figure}

This section investigates the dependence of spatially integrated heating
power, fast ion stored energy, and loss fraction on the RMP up-down phase
difference $\mathLaplace \Phi$. The results are shown in Fig.
\ref{21-5-26-a1}.

The first observation from Fig. \ref{21-5-26-a1} is that the heating power and
fast ion stored energy are reduced relative to the no RMP case and the loss
fraction is increased relative to the no RMP case. Fig. \ref{21-5-26-a1} shows
that the results for full vacuum RMP and the filtered vacuum RMP (keeping only
the $n = \pm 1$ Fourier harmonic) agree with each other well, indicating
harmonics other than $n = \pm 1$ have negligible effects. The results also
indicate that the heating powers and fast ion stored energy are roughly
decreasing functions of the loss fraction in all the three cases, as is
expected.

As is mentioned above, plasma response usually helps heal magnetic islands
and thus is believed to be beneficial to fast ion confinement. Comparing
between the response RMP case and the vacuum RMPs, however, we found the loss
fraction in the response RMP case is not always smaller the corresponding
vacuum case. This may indicate that non-resonant components, which can be
enhanced when plasma response is included, have significant effects on fast
ion confinement.

\begin{figure}[htp]
  \includegraphics[scale=0.6]{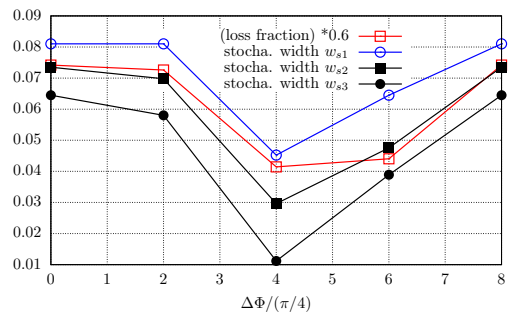}
  \caption{ \label{21-5-5-p1}Radial width of stochastic magnetic field as a
  function of the RMP coil up-down phase difference $\Delta \Phi$. The width
  is defined in terms of the square root of the normalized poloidal magnetic
  flux, $\rho = \sqrt{\overline{\Psi}_p}$ and is defined as the difference of
  $\rho$ between the stochastic radial location and the unperturbed
  last-closed-flux-surface. Three possible radial width, namely $w_{s 1}$,
  $w_{s 2}$, $w_{s 3}$, are plotted. Also plotted is the dependence of the
  loss fraction on the phase difference $\Delta \Phi$.}
\end{figure}

To more accurately describe the magnetic stochasticity, we use a purely
numerical method to determine the radial width of stochasticity (rather than
using the approximate analytical formula of magnetic island width and the
Chirkikov magnetic island overlap criteria discussed above). Figure
\ref{21-5-5-p1} plots the radial width of stochastic magnetic field as a
function of the RMP coil up-down phase difference $\Delta \Phi$. The radial
width is determined numerically by tracing a series of field lines starting
from the low field side midplane (i.e., $\theta = 0$, $\phi = [0 : 2 \pi]$,
and $\rho_p = [0 : 1]$). If a field line touches the wall during its first
3000 toroidal transits, then the magnetic field is considered as stochastic at
the initial radial location. At each toroidal location, scanning the initial
points from the core to the edge, the radial innermost location where magnetic
field becomes stochastic determines the width of the stochastic region. This
will give different value of radial width at different toroidal locations. The
reasonable radial width of stochasticity should be the toroidal maximal
values. This width is denoted by $w_{s 1}$ in Fig. \ref{21-5-5-p1}. As a
comparison, the toroidal minimal values and toroidal average values of the
width of stochasticity, denoted by $w_{s 2}$ and $w_{s 3}$, respectively are
also shown in Fig. \ref{21-5-5-p1}. The results show that the dependence of
fast ion loss fraction on $\Delta \Phi$ roughly agrees with those of the
radial width of stochasticity on $\Delta \Phi$.

\section{Resonance between RMPs and lost fast ions}\label{21-6-2a5}

Fast ion loss induced by RMP is often related to the resonance between fast
ions and RMPs. Generally, the resonance condition in tokamaks between a wave
and a particle drift is given by
\begin{equation}
  \label{3-8-e4m1} n \omega_{\phi} - \omega = p \omega_{\theta},
\end{equation}
where $n$ and $\omega$ are the toroidal mode number and angular frequency of
the wave, respectively, $\omega_{\phi}$ and $\omega_{\theta}$ are the toroidal
angular frequency and poloidal angular frequency of the particle drift,
respectively, and $p$ can be called resonance order, which is an arbitrary
rational number. When $p$ is an integer, it is the usual resonance condition.
When $p = j / k$ is a non-integer fraction ($j$ and $k$ are integers), the
resonance is called fractional resonance
{\cite{heidbrink2021,chen2019,kramer2012,Todo2018}} , where the (unperturbed)
drift motion reach the same phase of the wave when it finish $k$ times of
poloidal periods. The fractional resonance is often considered to be important
when the wave amplitude is large and thus might be related to the nonlinear
effects. However, the definition of resonance given above is essentially
linear because we are using the unperturbed motion in defining $\omega_{\phi}$
and $\omega_{\theta}$.

For RMPs with $\omega = 0$ and $n = 1$ considered in this work, the resonance
condition in Eq. (\ref{3-8-e4m1}) is simplified as
\begin{equation}
  \label{21-10-25-p1} p = \frac{\omega_{\phi}}{\omega_{\theta}} .
\end{equation}

To show how important the resonant effects are in generating fast ion loss, we
propose to use the graph of $\omega_{\phi} / \omega_{\theta}$ versus $\Delta
f$, where $\Delta f = f_{\tmop{rmp}} - f_{\tmop{normp}}$ is the difference of
the lost fast ion distribution function between the case with RMP and that
without RMP. By taking the difference, we exclude all the fast ions that are
lost due to the first-orbit prompt loss and pure collision loss.

\begin{figure}[htp]
  \includegraphics[scale=0.8]{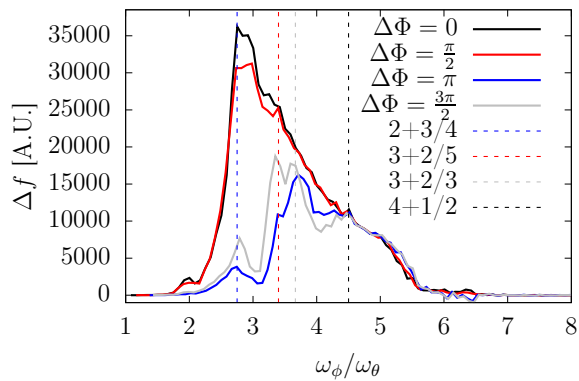}
  \caption{\label{21-4-13-a1}The difference of distribution of lost passing
  particles relative to the no RMP case for various RMP up-down phasings
  $\Delta \Phi$. The results show that some peaks of $\Delta f$ appear at
  non-integer values of $p$. The small peak at $p = 2$ is probably due to
  resonance of integer order. The peaks at $p = 2 + 3 / 4$ can be roughly
  regarded to be due to resonance of integer order, considering that the peaks
  are of finite width and are close to $p = 3$. The peaks at $p = 3 + 2 / 5$,
  $p = 2 + 2 / 3$ and $p = 4 + 1 / 2$ can not be well approximated as near
  integer values of $p$.}
\end{figure}

For the tangential NBI considered in this work, the results indicate that lost
particles are dominated by passing particles and we will consider only passing
particles here. The graph of $\omega_{\phi} / \omega_{\theta}$ versus $\Delta
f$ for lost passing ions are plotted in Fig. \ref{21-4-13-a1} for RMPs of
$\Delta \Phi = 0, \pi / 2, \pi, 3 \pi / 2$. The values of $\omega_{\phi} /
\omega_{\theta}$ at peaks of $\Delta f$ are indicated in Fig.
\ref{21-4-13-a1}, which shows that some of them can not be well approximated
by integers.

\

\begin{figure}[htp]
  \includegraphics[scale=0.8]{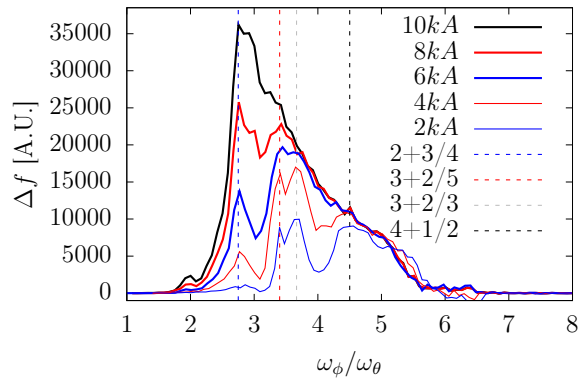}
  \caption{\label{21-6-1a2}The same as Fig. \ref{21-4-13-a1} except that the
  RMP up-down phasing is fixed at $\Delta \Phi = 0$ and the amplitude of the
  coil current is scanned.}
\end{figure}

As is indicated in Fig. \ref{21-4-13-a1}, some peaks of $\Delta f$ seem to
appear at non-integer fractions. It is usually believed that fractional
resonances are associated with nonlinear effects and thus amplitude of the
perturbation matters. In the work of Kramer et al.{\cite{kramer2012}}, the
fractional resonance appears when the amplitude of the wave exceeds a
threshold and thus it is natural to connect the fraction resonance with
nonlinear effects. Here we examine how peaks of $\Delta f$ change with the
changing of amplitude of coil currents. The results are plotted in Fig.
\ref{21-6-1a2}, where the peaks remain at the same non-integer values when the
current amplitude is changed from $10 \tmop{kA}$ (the designed maximal total
current in RMP coils on EAST) to $2 \tmop{kA}$. The fact that these peaks
persist even when the RMP amplitude is very small (2kA) may imply that the
peaks at these non-integer values of $\omega_{\phi} / \omega_{\theta}$ are not
related to nonlinear effects.

We note that linear resonances can be of non-integer fractional orders (it is
easy to demonstrate this by deriving the resonance condition from
scratch{\cite{Todo2018}}). Since, as mentioned above, it is hard to relate the
peaks at non-integer fraction with nonlinear effects, a plausible
interpretation for these peaks is that they are due to linear resonances of
non-integer fractional orders.

We note that resonance is not a necessary condition for RMP to induce fast ion
loss (resonance is only a beneficial case for perturbation to have large
impact). Any perturbation that breaks the axisymmetry can potentially generate
radial particle transport and hence loss. Therefore, we can not exclude the
possibility that the peaks at non-integer values are caused by non-resonant
effects of RMPs on fast ions. This is one of the limitations of this crude
statistical method in identifying resonance.

Another observation from Fig. \ref{21-6-1a2} is that the peaks are of nonzero
expansion width, so that the consecutive peaks are not well separated, which
makes it difficult to identify all peaks in some cases (e.g., for the case of
10kA RMP, only the peaks near $p = 2, 3$ show up and all the other peaks merge
and are blurred).

\begin{figure}[htp]
  \includegraphics[scale=0.5]{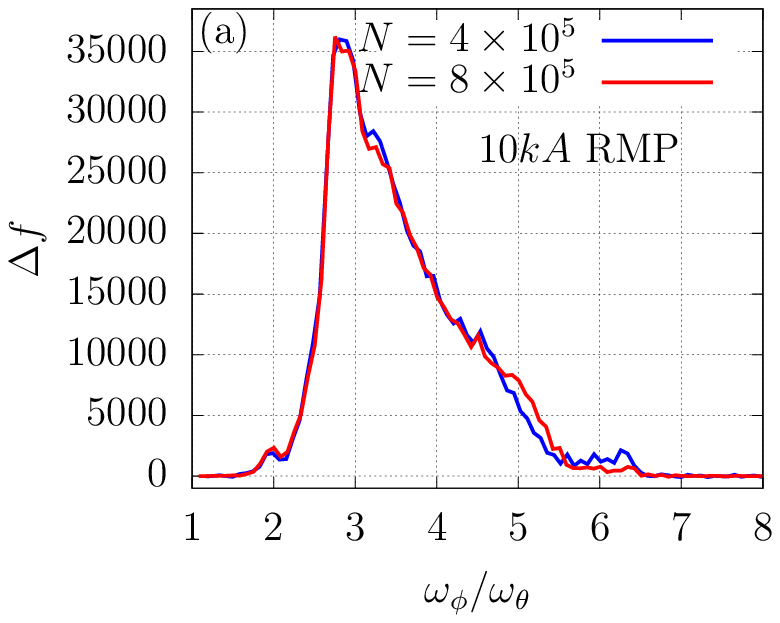}\includegraphics[scale=0.5]{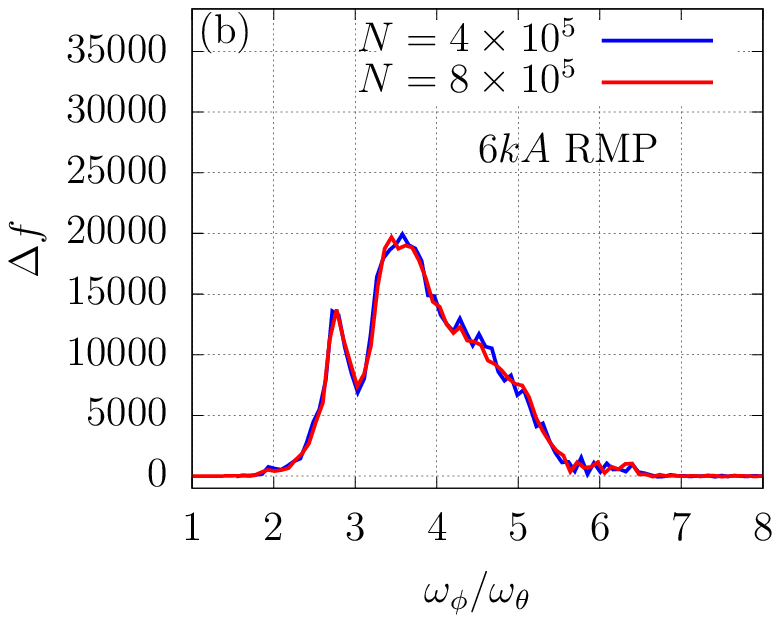}
  \caption{\label{21-6-1a3}Convergence of $\Delta f$ over number of markers
  $N$ for RMPs of currents $10 \tmop{kAt}$ (left) and $6 \tmop{kAt}$ (right).
  RMPs of $\Delta \Phi = 0$ are used in the simulation.}
\end{figure}

Since there are obvious statistic noises in the results of Figs.
\ref{21-4-13-a1} and \ref{21-6-1a2}, we need to make sure that the dominant
peaks identified above are not sensitive to the noise. We perform a numerical
convergence study over the number of markers $N$ used in the simulation (which
is the most important numerical parameter in reducing noise). The results are
plotted in Fig. \ref{21-6-1a3}, which shows that the statistic noise does no
significantly change the peaks in $\Delta f$, i.e., the results are well
converged, in terms of the dominant resonant peaks.

\begin{figure}[htp]
  \includegraphics[scale=0.8]{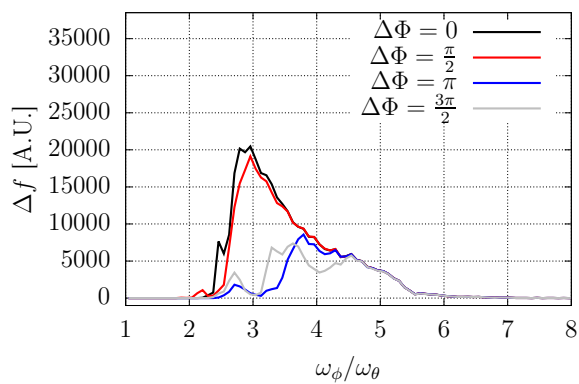}
  \caption{\label{21-11-2-p1}The same as Fig. \ \ref{21-4-13-a1} except that
  collisions are turned off in this case. The peaks of $\Delta f$ roughly
  agree with those in Fig. \ref{21-4-13-a1}. The results also show that the
  amplitude of $\Delta f$ in the collisionless case is much smaller than that
  in the collisional case (Fig. \ \ref{21-4-13-a1}). The unit of $\mathLaplace
  f$ is the same as that of Fig. \ \ref{21-4-13-a1}. }
\end{figure}

In the above, $\Delta f$ is the difference of the lost fast ion distribution
function between the case with RMP and that without RMP. Collisions are
included in both the cases. By taking the difference, we expect to cancel the
collision effect and obtain the pure RMP effect. However collision may blur
the resonance between RMPs and fast ions if the collision effects are not
exactly canceled out in the simple subtraction. Thus it is desirable to do a
simulation with collisions turned off. The results are plotted in Fig.
\ref{21-11-2-p1} for RMPs of $\Delta \Phi = 0, \pi / 2, \pi, 3 \pi / 2$, which
show similar behavior as that in Fig. \ \ref{21-4-13-a1}. Specifically, there
are peaks at non-integer values of $p$ (e.g., peaks between $p = 3$ and $p =
4$ for the $\Delta \Phi = \pi$ and $\Delta \Phi = 3 \pi / 2$ RMPs). There are
also some peaks which are very near to integer values of $p$ (e.g., the peak
near $p = 3$ for the $\Delta \Phi = \pi / 2$ RMP).

\section{Summary and discussion}\label{19-12-26-8}\label{19-9-18-1}

The effects of resonant magnetic perturbation on the steady-state radial
profile of neutral beam heating and fast ion pressure are studied
computationally in a realistic tokamak configuration. It is observed in the
simulations that RMPs can accumulate fast ions at some radial locations,
giving a higher fast ion pressure than the case without RMP. RMPs generally
have pump out effects on fast ions. The pump out effects are non-uniform along
the radial direction, which implies that some radial locations may accumulate
fast ions. Therefore it is not surprising to observe the local increasing of
fast ion pressure when a RMP is imposed.

It is also found that the toroidal phasing of the RMP with respect to the fast
ion source has slight effects on the steady-state radial profile of fast ions.
The effect being slight is expected, considering that most particles resulted
from the tangential injection are passing particles which quickly traverse a
full toroidal loop, making the relative location between neutral beam source
and RMP coils almost irrelevant. For trapped particles resulted from
perpendicular NBI whose toroidal procession frequencies are low, the toroidal
phase of RMP coils relative to the source can be important since the trapped
particles can remain in a limited toroidal range for a sufficient time. The
toroidal phasing can be a useful free parameter for fast ion control using
RMPs, considering that ELMs control is not sensitive to the toroidal phasing.

The dependence of fast ion loss fraction on the RMP up-down phase difference
is found to show similar behavior as the dependence of the radial width of
chaotic edge magnetic field on the up-down phase difference.

A simple numerical statistical method, which uses the lost fast ion
distribution over $\omega_{\phi} / \omega_{\theta}$, is proposed to identify
resonance between lost fast ions and RMPs. Using this method, we found that
the RMP induced loss of passing particles may be partially due to resonance of
fractional orders. However, the reliability and usefulness of this statistic
method in identifying resonance between lost fast ions and RMPs needs to be
further explored. Most authors analyze fast ion transport in terms of the
variation of the toroidal canonical momentum, $\delta P_{\phi}$, which is a
more informative figure of merit than the simple condition of loss or not
(used in this work), in revealing the resonance between fast-ions and magnetic
perturbations{\cite{Sanchis_2018}}.

\section{ACKNOWLEDGMENTS}

One of the authors (Y. Hu) acknowledges useful discussions with Dr. Nong Xiang
and Dr. Wei Chen. Numerical computations were performed on Tianhe at National
SuperComputer Center in Tianjin and the ShenMa computing cluster in Institute
of Plasma Physics, Chinese Academy of Sciences. This work was supported by
National Key R\&D Program of China under Grant No. 2017YFE0300400, by
Comprehensive Research Facility for Fusion Technology Program of China under
Contract No. 2018-000052-73-01-001228, by users with Excellence Program of
Hefei Science Center CAS under Grant No. 2021HSC-UE017, and by the National
Natural Science Foundation of China under Grant No. 11575251.

\subsection{Conflict of interest}

The authors have no conflicts to disclose.

\section{DATA AVAILABILITY}

The data that support the findings of this study are available from the
corresponding author upon reasonable request.

\appendix\section{Neutral particle source}\label{21-4-14-p1}

\begin{figure}[htp]
  \includegraphics[scale=0.9]{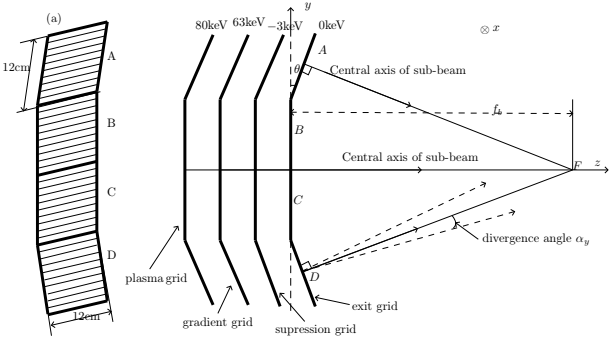}
  \caption{\label{17-2-17-1}Left: Three-dimensional sketch map of one of the
  accelerating grids. Right: Side view of the four groups of accelerating
  grids of EAST neutral beam injector, which are, respectively, called plasma
  grid, gradient grid, suppression grid, and exit grid. A typical setting of
  voltage on the grids is indicated on the graph. Each accelerating grid has
  four sub-grids, indicated by $A$, $B$, $C$, and $D$ on the figure. Each
  sub-grid is a $12 \tmop{cm} \times 12 \tmop{cm}$ square. Sub-grids $A$ and
  $D$ are rotated with respect to the central sub-grids $B$ and $C$ by a small
  angle $\theta = \left( 1 \frac{1}{12} \right)^{\circ}$. This angle is
  exaggerated on the figure. The central axis of the two beams from grids $A$
  and $D$ intersect at $F$. The vertical focal length $f_b$ is defined as the
  distance from point $F$ to the plane of $B C$ grids ($f_b \approx 9.5 m$).
  The horizontal focal length is infinite. The $y = 0$ plane corresponds to
  the tokamak midplane. The beam horizontal and vertical divergence angles are
  set to be $0.6^{\circ}$ and $1.2^{\circ}$, respectively.}
\end{figure}

Neutral particle source is implemented by using Monte-Carlo method and taking
into account the beam focus and divergence, the shape of the accelerating
grids, and the spatial distribution of particles on the exit grid. Figure
\ref{17-2-17-1} shows a sketch map of the accelerating grids of EAST neutral
beam system.

The neutralizing process is not directly modeled in the simulation and its
effect is included only through the input number ratio between particles of
the full (kinetic) energy, half energy and $1 / 3$ energy (the number ratio
between them is assumed to be $80 : 14 : 6$ in this paper, with the full
energy being $51 \tmop{keV}$). We assume that other beam properties (e.g.
focus and divergence) are not changed by the neutralizing process, except that
only part of the ions succeed in becoming neutrals. The NBI power after
neutralizing is assumed to be 1MW. In this paper, we consider one of the four
beams on EAST, that is injected in the co-current direction, with the tangent
radius of the central beam being $1.26 m$.

\section{Ionization in tokamak plasma}\label{21-5-30-a3}\label{21-4-14-p2}

The neutral particle trajectories are assumed to be along straight lines until
they arrive at the ionization locations in the plasma. The Monte-Carlo method
of implementing the ionization process is as follows{\cite{pankin2004}}. First
define $\nu$ by
\begin{equation}
  \nu = n_i \sigma_{\tmop{ch}} + n_i \sigma_i + n_e \frac{\langle \sigma_e v_e
  \rangle}{v_b},
\end{equation}
where $n_i$ and $n_e$ are the number density of background plasma ions and
electrons, \ respectively, $\sigma_{\tmop{ch}}$ is the cross-section for
charge exchange with plasma ions, $\sigma_i$ are the cross-section for
ionization by plasma ions, $\langle \sigma_e v_e \rangle$ is the electron
impact ionization rate coefficient averaged over the Maxwellian distribution,
$\langle \sigma_e v_e \rangle / v_b$ is the effective cross-section of
electron impact ionization, where $v_b$ is the neutral particle velocity. Then
associate each marker loaded with a random number $\eta$ that is uniformly
distributed in $[0, 1]$. Then, along the trajectory of each neutral particle
(straight line), the integration $s = \int_0^l \nu (l') d l'$ is calculated to
examine whether $s \geqslant \ln (1 / \eta)$ or not. If $s \geqslant \ln (1 /
\eta)$, the neutral particle is considered to be ionized. The value of $\nu$
outside the last closed flux surface (LCFS) is set to be zero, i.e., the
ionization outside LCFS is not considered. The ionization cross-sections data
used in this work are from the ADAS database (https://open.adas.ac.uk/) and
Janev's paper{\cite{janev1993}}.

The Monte-Carlo implementation gives the ionization locations of each marker
loaded in the simulation. These locations are used as initial conditions in
the subsequent orbit following computations. Those neutral particles that are
not yet ionized when they reach the inner wall of the device are lost to the
wall and this loss are called shine-through loss.

\section{Guiding-center motion and finite Larmor radius effect}

Knowing the birth location $\mathbf{x}$ of a fast ion (given by the module
calculating the neutral particle ionization), the corresponding guiding-center
location $\mathbf{X}$ is calculated via the following guiding-center
transform:
\begin{equation}
  \label{16-9-21-1} \mathbf{X}=\mathbf{x}+\mathbf{v} \times \frac{\mathbf{b}
  (\mathbf{x})}{\Omega (\mathbf{x})},
\end{equation}
where $\mathbf{v}$ is the fast ion velocity, $\mathbf{b}=\mathbf{B}/ B$, \
$\Omega = B Z_f e / m_f$ is the cyclotron angular frequency, $m_f$ and $Z_f e$
are the mass and charge of the fast ion, respectively, $\mathbf{B}$ is the
magnetic field.

The guiding-center drift of each fast ion is then followed by numerically
integrating the following guiding center motion equation{\cite{todo2006}}:
\begin{equation}
  \label{5-15-p2} \frac{d\mathbf{X}}{d t} =
  \frac{\mathbf{B}^{\star}}{B^{\star}_{\parallel}} v_{\parallel} +
  \frac{\mu}{m_f \Omega B^{\star}_{\parallel}} \mathbf{B} \times \nabla B,
\end{equation}
\begin{equation}
  \label{16-7-4-7} \frac{d v_{\parallel}}{d t} = - \frac{\mu}{m_f } 
  \frac{\mathbf{B}^{\star}}{B^{\star}_{\parallel}} \cdot \nabla B
\end{equation}
where $v_{\parallel}$ is the parallel (to the magnetic field) velocity, $\mu$
is the magnetic moment (a constant of motion) defined by $\mu = m_f
v_{\perp}^2 / (2 B)$ with $v_{\perp}$ being the perpendicular speed; \
$\mathbf{B}^{\star}$ and $B^{\star}_{\parallel}$ are defined by
\begin{equation}
  \mathbf{B}^{\star} =\mathbf{B}+ B \frac{v_{\parallel}}{\Omega} \nabla \times
  \mathbf{b},
\end{equation}

\begin{equation}
  \label{5-15-p8} B^{\star}_{\parallel} \equiv \mathbf{b} \cdot
  \mathbf{B}^{\star} = B \left( 1 + \frac{v_{\parallel}}{\Omega} \mathbf{b}
  \cdot \nabla \times \mathbf{b} \right),
\end{equation}
respectively.

The cylindrical coordinates $(R, \phi, z)$ are adopted in writing the
component equations of guiding-center motion. Using cylindrical coordinates
(rather than magnetic coordinates) has the advantage that we can handle orbits
on the magnetic axis and outside the LCFS without difficulties. The 4th order
Runge-Kutta scheme is used in integrating the equations. Orbits outside the
LCFS are followed until they touch the wall.

The finite Larmor radius (FLR) effect is taken into account by evaluating and
averaging the magnetic fields in Eqs. (\ref{5-15-p2}) and (\ref{16-7-4-7}) on
a gyro-ring (four points average is used in the simulation). The gyro-ring is
approximated by a circle in the poloidal plane with the Larmor radius
calculated by using the toroidal magnetic field at the guiding-center
location. When checking whether a fast ion touches the wall, four points on
the gyro-ring are also calculated. If any one of the four points touches the
wall, the fast ion is considered as lost.

\begin{figure}[htp]
  \includegraphics{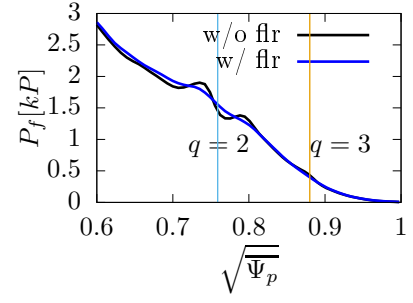}
  \caption{\label{21-6-1p1}Comparison of radial profiles of fast ion pressure
  between the case when the FLR effect is taken into account and that when it
  is neglected. The FLR effect is taken into account when checking whether
  particles touch the first wall in both the cases. RMP of $\Delta \Phi = 0$
  is used.}
\end{figure}

The FLR effect is also taken into account when depositing markers to compute
fast ion pressure. Neglecting the FLR effect in this case will give
significantly different radial profile of fast ion steady-state pressure, as
is show in Fig. \ref{21-6-1p1}, which indicates that there is significant
averaging effect from the large Larmor radius of fast ions.

\section{Fast ion collisions and heating power density}\label{21-5-17-a1}

The collision model of fast ions with the background electrons and ions
adopted in {\codestar{TGCO}} includes the effects of the slowing-down, energy
diffusion, and pitch angle scattering.

The birth velocity of the fast ions from NBI is much larger than the thermal
velocity of background ions but still much smaller than the electron thermal
velocity, i.e,
\begin{equation}
  \label{9-10-1} v_{t i} \ll v_f \ll v_{t e} .
\end{equation}
(This condition is still valid for the fast alpha particles in reactor
plasmas.) Using this condition, the collision term of fast ions with the
background Maxwellian electrons and ions can be simplified. Specifically, the
Monte-Carlo implementation of the resulting collision operator takes the form
given in Ref. {\cite{boozer1981,todo2014}}. For reference ease, we repeat it
here. In the Monte-Carlo implementation, the pitch-angle variable $\lambda =
v_{\parallel} / v$ and velocity $v$ are altered at the end of each time step
according to the following scheme:
\begin{equation}
  \label{19-12-26-1} \lambda_{\tmop{new}} = \lambda_{\tmop{old}} (1 - \nu_d
  \Delta t) \pm \sqrt{(1 - \lambda_{\tmop{old}}^2) \nu_d \Delta t},
\end{equation}
and
\begin{eqnarray}
  v_{\tmop{new}} & = & v_{\tmop{old}} - v_{\tmop{old}} \nu_s \Delta t \left[ 1
  + \left( \frac{v_c}{v_{\tmop{old}}} \right)^3 \right] \nonumber\\
  & + & \frac{\nu_s \Delta t}{m_f v_{\tmop{old}}} \left[ T_e - \frac{1}{2}
  T_i \left( \frac{v_c}{v_{\tmop{old}}} \right)^3 \right] \nonumber\\
  & \pm & \sqrt{\frac{\nu_s \Delta t}{m_f} \left[ T_e + T_i \left(
  \frac{v_c}{v_{\tmop{old}}} \right)^3 \right]}  \label{19-12-26-2}
\end{eqnarray}

where $\pm$ is randomly chosen with equal probability for plus and minus,
$\Delta t$ is the time step, $\nu_d$ is the velocity-dependent pitch-angle
scattering rate given by
\begin{equation}
  \nu_d = \frac{Z_{\tmop{eff}}}{v^3} \Gamma^{f / e} .
\end{equation}
with $\Gamma^{f / e}$ defined by
\begin{equation}
  \Gamma^{f / e} = \frac{n_e Z_f^2 e^4}{4 \pi \epsilon_0^2 m_f^2} \ln
  \Lambda^{f / e},
\end{equation}
where $Z_{\tmop{eff}}$ is the effective charge number of background ions
($Z_{\tmop{eff}} = 1$ in this work since we assume a pure Deuterium plasma
without impurities), $Z_f$ is the fast ion charge number ($Z_f = 1$ in this
work). $\epsilon_0$ and $\ln \Lambda^{f / e}$ are, respectively, vacuum
dielectric constant and the Coulomb logarithm. In Eq. (\ref{19-12-26-2}),
$\nu_s$ is the fast ion slowing down rate due to the background electrons and
is given by
\begin{equation}
  \label{20-1-17-a1} \nu_s = \frac{4}{3 \sqrt{\pi}} \frac{m_f}{m_e} 
  \frac{\Gamma^{f / e}}{\left( \sqrt{2 T_e / m_e} \right)^3} .
\end{equation}
In Eq. (\ref{19-12-26-2}), $v_c$ is the critical velocity (at which the
friction due to background ions is equal to that due to background electrons)
given by
\begin{equation}
  \label{19-12-24-1} v_c = \left(  \frac{3 \sqrt{\pi}}{4} \frac{m_e}{m_i}
  \right)^{1 / 3} \sqrt{\frac{2 T_e}{m_e}} .
\end{equation}
The scheme in Eq. (\ref{19-12-26-1}) models the pitch-angle scattering (due to
background ions only). The first line of Eq. (\ref{19-12-26-2}) models the
slowing-down (due to both background electrons (the ``1'' term) and ions (the
$(v_c / v_{\tmop{old}})^3$ term)) and the second line models the energy
diffusion (due to both background ions and electrons). Collisions between fast
ions themselves are ignored.

\begin{figure}[htp]
  \includegraphics[scale=0.9]{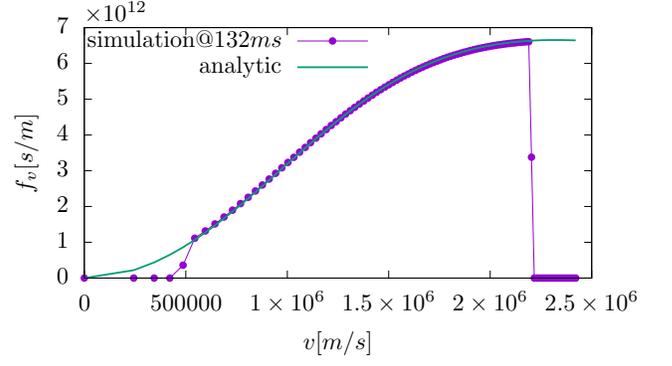}
  \caption{\label{19-12-25-p1}Comparison between the numerical steady state
  velocity distribution function and the analytic slowing-down distribution
  function. The distribution function is defined by $f_v d v = d N$ with $d N$
  being the number of particles with velocity in $[v, v + d v]$. There is a
  discrepancy in the small velocity region because we consider particles of
  kinetic energy less than $2 T_{i 0}$ as thermalized and do not include their
  contribution to the fast ion distribution. There is a numerical jump at the
  injection energy since no particles are accelerated to higher energy (energy
  diffusion could make this jump smoother if included).}
\end{figure}

As a simple verification of the implementation of the numerical slowing-down
model, Fig. \ref{19-12-25-p1} compares the numerical steady state velocity
distribution function and the analytic slowing-down distribution function,
which shows good agreement between them. The analytic slowing-down
distribution is given by
\begin{equation}
  f_v \propto \frac{v^2}{(1 + (v / v_c)^3)},
\end{equation}
where $v_c$ is the critical velocity given by Eq. (\ref{19-12-24-1}). The
numerical result was obtained in a simplified setting by turning off orbiting
(hence no edge loss) and assuming uniform plasma profile, no energy diffusion,
and injection with single energy of $51 \tmop{keV}$ in EAST discharge
\#52340@3.4s.

The heating power (due to fast ions) to bulk plasma within a given spatial
volume $V$ is equal to the fast ion kinetic energy loss rate within that
volume, i.e.,
\begin{equation}
  \label{20-4-1-p1} P = \sum_{j = 1}^N w_j \frac{\Delta E_j}{\Delta t},
\end{equation}
where $N$ is the number of markers within the given spatial volume, $w_j$ is
the marker weight, $\Delta E_j$ is the kinetic energy decrease of a marker
during the time interval $\Delta t$. Using $\Delta E_j = \frac{1}{2} m_f v_{j,
\tmop{old}}^2 - \frac{1}{2} m_f v_{j, \tmop{new}}^2$, and the slowing-down
algorithm, expression (\ref{20-4-1-p1}) is written as
\begin{eqnarray*}
  &  & P = \frac{1}{2} m_f \sum_{j = 1}^N\\
  &  & w_j [2 v_{j, \tmop{old}}^2 \nu_s (C_i + C_e) - v_{j, \tmop{old}}^2
  \nu_s (C_i + C_e)^2 \nu_s \Delta t]
\end{eqnarray*}
where $C_i = (v_c / v_{j, \tmop{old}})^3, C_e = 1$. Neglecting the second
term, which is smaller than the first term by a factor of $\nu_s \Delta t \ll
1$, the above expression is written as
\begin{eqnarray}
  P & \approx & \frac{1}{2} m_f \sum_{j = 1}^N w_j [2 v_{j, \tmop{old}}^2
  \nu_s (C_i + C_e)] \nonumber\\
  & = & P_i + P_e, 
\end{eqnarray}
where
\begin{equation}
  P_i = m_f \sum_{j = 1}^N w_j v_{j, \tmop{old}}^2 \nu_s C_i,
\end{equation}
and
\begin{equation}
  P_e = m_f \sum_{j = 1}^N w_j v_{j, \tmop{old}}^2 \nu_s C_e,
\end{equation}
which are the heating power delivered to the thermal ions and electrons,
respectively.

The heating power densities are then given by ${H_i}  = P_i / V$ and ${H_e}  =
P_e / V$.

\nocite{*}

\begin{thebibliography}{10}
  \bibitem[1]{Scoville2019}J.~Scoville, M.~Boyer, B.~Crowley, N.~Eidietis,
  C.~Pawley, and J.~Rauch, {\newblock}Fusion Engineering and Design
  \tmtextbf{146}, 6 (2019), {\newblock}SI:SOFT-30.
  
  \bibitem[2]{Hu2015}C.~Hu, Y.~Xie, Y.~Xie, S.~Liu, Y.~Xu, L.~Liang, C.~Jiang,
  P.~Sheng, Y.~Gu, J.~Li, and Z.~Liu, {\newblock}Plasma Science and Technology
  \tmtextbf{17}, 817 (2015).
  
  \bibitem[3]{Schneider_2011}M.~Schneider, L.-G. Eriksson, I.~Jenkins,
  J.~Artaud, V.~Basiuk, F.~Imbeaux, T.~Oikawa, and and, {\newblock}Nuclear
  Fusion \tmtextbf{51}, 063019 (2011).
  
  \bibitem[4]{ASUNTA201533}O.~Asunta, J.~Govenius, R.~Budny, M.~Gorelenkova,
  G.~Tardini, T.~Kurki-Suonio, A.~Salmi, and S.~Sipil{\"a},
  {\newblock}Computer Physics Communications \tmtextbf{188}, 33 (2015).
  
  \bibitem[5]{Heidbrink_2009}W.~W. Heidbrink, M.~Murakami, J.~M. Park, C.~C.
  Petty, M.~A.~V. Zeeland, J.~H. Yu, and G.~R. McKee, {\newblock}Plasma
  Physics and Controlled Fusion \tmtextbf{51}, 125001 (2009).
  
  \bibitem[6]{pankin2004}A.~Pankin, D.~McCune, R.~Andre, G.~Bateman, and
  A.~Kritz, {\newblock}Computer Physics Communications \tmtextbf{159}, 157
  (2004).
  
  \bibitem[7]{Wan_2017}B.~Wan, Y.~Liang, X.~Gong, J.~Li, N.~Xiang, G.~Xu,
  Y.~Sun, L.~Wang, J.~Qian, H.~Liu, X.~Zhang, L.~Hu, J.~Hu, F.~Liu, C.~Hu,
  Y.~Zhao, L.~Zeng, M.~Wang, H.~Xu, G.~Luo, A.~Garofalo, A.~Ekedahl, L.~Zhang,
  X.~Zhang, J.~Huang, B.~Ding, Q.~Zang, M.~Li, F.~Ding, S.~Ding, B.~Lyu,
  Y.~Yu, T.~Zhang, Y.~Zhang, G.~Li, T.~Xia, and and, {\newblock}Nuclear Fusion
  \tmtextbf{57}, 102019 (2017).
  
  \bibitem[8]{Wan_2019}B.~Wan, Y.~Liang, X.~Gong, N.~Xiang, G.~Xu, Y.~Sun,
  L.~Wang, J.~Qian, H.~Liu, L.~Zeng, L.~Zhang, X.~Zhang, B.~Ding, Q.~Zang,
  B.~Lyu, A.~Garofalo, A.~Ekedahl, M.~Li, F.~Ding, S.~Ding, H.~Du, D.~Kong,
  Y.~Yu, Y.~Yang, Z.~Luo, J.~Huang, T.~Zhang, Y.~Zhang, G.~Li, T.~Xia, and
  and, {\newblock}Nuclear Fusion \tmtextbf{59}, 112003 (2019).
  
  \bibitem[9]{Sun_2016}Y.~Sun, M.~Jia, Q.~Zang, L.~Wang, Y.~Liang, Y.~Liu,
  X.~Yang, W.~Guo, S.~Gu, Y.~Li, B.~Lyu, H.~Zhao, Y.~Liu, T.~Zhang, G.~Li,
  J.~Qian, L.~Xu, N.~Chu, H.~Wang, T.~Shi, K.~He, D.~Chen, B.~Shen, X.~Gong,
  X.~Ji, S.~Wang, M.~Qi, Q.~Yuan, Z.~Sheng, G.~Gao, Y.~Song, P.~Fu, and B.~W.
  and, {\newblock}Nuclear Fusion \tmtextbf{57}, 036007 (2016).
  
  \bibitem[10]{sun2016prl}Y.~Sun, Y.~Liang, Y.~Q. Liu, S.~Gu, X.~Yang, W.~Guo,
  T.~Shi, M.~Jia, L.~Wang, B.~Lyu, C.~Zhou, A.~Liu, Q.~Zang, H.~Liu, N.~Chu,
  H.~H. Wang, T.~Zhang, J.~Qian, L.~Xu, K.~He, D.~Chen, B.~Shen, X.~Gong,
  X.~Ji, S.~Wang, M.~Qi, Y.~Song, Q.~Yuan, Z.~Sheng, G.~Gao, P.~Fu, and
  B.~Wan, {\newblock}Phys. Rev. Lett. \tmtextbf{117}, 115001 (2016).
  
  \bibitem[11]{Xu_2020}Y.~Xu, L.~Li, Y.~Hu, Y.~Liu, W.~Guo, L.~Ye, and
  X.~Xiao, {\newblock}Nuclear Fusion \tmtextbf{60}, 086013 (2020).
  
  \bibitem[12]{yxu2019}Y.~Xu, W.~Guo, Y.~Hu, L.~Ye, X.~Xiao, and S.~Wang,
  {\newblock}Computer Physics Communications \tmtextbf{244}, 40 (2019).
  
  \bibitem[13]{He_2019}K.~He, B.~Wan, Y.~Sun, M.~Jia, T.~Shi, H.~Wang, and
  X.~Zhang, {\newblock}Nuclear Fusion \tmtextbf{59}, 126026 (2019).
  
  \bibitem[14]{he2020a}K.~He, Y.~Sun, B.~Wan, S.~Gu, and M.~Jia,
  {\newblock}Nuclear Fusion \tmtextbf{60}, 126027 (2020).
  
  \bibitem[15]{he2020b}K.~He, Y.~Sun, B.~Wan, S.~Gu, M.~Jia, and Y.~Hu,
  {\newblock}Nuclear Fusion \tmtextbf{61}, 016009 (2020).
  
  \bibitem[16]{Garcia_Munoz_2013}M.~Garcia-Munoz, S.~{\"A}k{\"a}slompolo,
  P.~de~Marne, M.~G. Dunne, R.~Dux, T.~E. Evans, N.~M. Ferraro, S.~Fietz,
  C.~Fuchs, B.~Geiger, A.~Herrmann, M.~Hoelzl, B.~Kurzan, N.~Lazanyi, R.~M.
  McDermott, M.~Nocente, D.~C. Pace, M.~Rodriguez-Ramos, K.~Shinohara,
  E.~Strumberger, W.~Suttrop, M.~A.~V. Zeeland, E.~Viezzer, M.~Willensdorfer,
  and E.~Wolfrum, {\newblock}Plasma Physics and Controlled Fusion
  \tmtextbf{55}, 124014 (2013).
  
  \bibitem[17]{Garcia_Munoz_2013b}M.~Garcia-Munoz, S.~{\"A}k{\"a}slompolo,
  O.~Asunta, J.~Boom, X.~Chen, I.~Classen, R.~Dux, T.~Evans, S.~Fietz,
  R.~Fisher, C.~Fuchs, B.~Geiger, M.~Hoelzl, V.~Igochine, Y.~Jeon, J.~Kim,
  J.~Kim, B.~Kurzan, N.~Lazanyi, T.~Lunt, R.~McDermott, M.~Nocente, D.~Pace,
  T.~Rhodes, M.~Rodriguez-Ramos, K.~Shinohara, W.~Suttrop, M.~V. Zeeland,
  E.~Viezzer, M.~Willensdorfer, E.~Wolfrum, , and and, {\newblock}Nuclear
  Fusion \tmtextbf{53}, 123008 (2013).
  
  \bibitem[18]{Sanchis_2018}L.~Sanchis, M.~Garcia-Munoz, A.~Snicker, D.~A.
  Ryan, D.~Zarzoso, L.~Chen, J.~Galdon-Quiroga, M.~Nocente, J.~F.
  Rivero-Rodriguez, M.~Rodriguez-Ramos, W.~Suttrop, M.~A.~V. Zeeland,
  E.~Viezzer, M.~Willensdorfer, and F.~Z. and, {\newblock}Plasma Physics and
  Controlled Fusion \tmtextbf{61}, 014038 (2018).
  
  \bibitem[19]{kim2018}K.~Kim, H.~Jhang, J.~Kim, and T.~Rhee,
  {\newblock}Physics of Plasmas \tmtextbf{25}, 122511 (2018).
  
  \bibitem[20]{Van_Zeeland_2013}M.~A.~V. Zeeland, N.~M. Ferraro, W.~W.
  Heidbrink, G.~J. Kramer, D.~C. Pace, X.~Chen, T.~E. Evans, R.~K. Fisher,
  M.~Garc{\'i}a-Mu{\~n}oz, J.~M. Hanson, M.~J. Lanctot, L.~L. Lao, R.~A.
  Moyer, R.~Nazikian, and D.~M. Orlov, {\newblock}Plasma Physics and
  Controlled Fusion \tmtextbf{56}, 015009 (2013).
  
  \bibitem[21]{Van_Zeeland_2015}M.~V. Zeeland, N.~Ferraro, B.~Grierson,
  W.~Heidbrink, G.~Kramer, C.~Lasnier, D.~Pace, S.~Allen, X.~Chen, T.~Evans,
  M.~Garc{\'i}a-Mu{\~n}oz, J.~Hanson, M.~Lanctot, L.~Lao, W.~Meyer, R.~Moyer,
  R.~Nazikian, D.~Orlov, C.~Paz-Soldan, and A.~Wingen, {\newblock}Nuclear
  Fusion \tmtextbf{55}, 073028 (2015).
  
  \bibitem[22]{McClements_2015}K.~G. McClements, R.~J. Akers, W.~U. Boeglin,
  M.~Cecconello, D.~Keeling, O.~M. Jones, A.~Kirk, I.~Klimek, R.~V. Perez,
  K.~Shinohara, and K.~Tani, {\newblock}Plasma Physics and Controlled Fusion
  \tmtextbf{57}, 075003 (2015).
  
  \bibitem[23]{Tani1981}K.~Tani, M.~Azumi, H.~Kishimoto, and S.~Tamura,
  {\newblock}Journal of the Physical Society of Japan \tmtextbf{50}, 1726
  (1981).
  
  \bibitem[24]{HIRVIJOKI2014}E.~Hirvijoki, O.~Asunta, T.~Koskela,
  T.~Kurki-Suonio, J.~Miettunen, S.~Sipil{\"a}, A.~Snicker, and
  S.~{\"A}k{\"a}slompolo, {\newblock}Computer Physics Communications
  \tmtextbf{185}, 1310 (2014).
  
  \bibitem[25]{White_2010}R.~B. White, N.~Gorelenkov, W.~W. Heidbrink, and
  M.~A.~V. Zeeland, {\newblock}Plasma Physics and Controlled Fusion
  \tmtextbf{52}, 045012 (2010).
  
  \bibitem[26]{Kramer_2013}G.~J. Kramer, R.~V. Budny, A.~Bortolon, E.~D.
  Fredrickson, G.~Y. Fu, W.~W. Heidbrink, R.~Nazikian, E.~Valeo, and M.~A.~V.
  Zeeland, {\newblock}Plasma Physics and Controlled Fusion \tmtextbf{55},
  025013 (2013).
  
  \bibitem[27]{PFEFFERLE20143127}D.~Pfefferl{\'e}, W.~Cooper, J.~Graves, and
  C.~Misev, {\newblock}Computer Physics Communications \tmtextbf{185}, 3127
  (2014).
  
  \bibitem[28]{xyXU_2020}X.~XU, Y.~XU, X.~ZHANG, Y.~HU, L.~YE, and X.~XIAO,
  {\newblock}Plasma Science and Technology \tmtextbf{22}, 085101 (2020).
  
  \bibitem[29]{liu2000}Y.~Q. Liu, A.~Bondeson, C.~M. Fransson, B.~Lennartson,
  and C.~Breitholtz, {\newblock}Physics of Plasmas \tmtextbf{7}, 3681 (2000).
  
  \bibitem[30]{liu2010}Y.~Liu, A.~Kirk, and E.~Nardon, {\newblock}Physics of
  Plasmas \tmtextbf{17}, 122502 (2010).
  
  \bibitem[31]{cheng1987}C.~Cheng and M.~Chance, {\newblock}J. of Comput.
  Phys. \tmtextbf{71}, 124 (1987).
  
  \bibitem[32]{boozer2001prl}A.~H. Boozer, {\newblock}Phys. Rev. Lett.
  \tmtextbf{86}, 5059 (2001).
  
  \bibitem[33]{Sun_2015}Y.~Sun, Y.~Liang, J.~Qian, B.~Shen, and B.~Wan,
  {\newblock}Plasma Physics and Controlled Fusion \tmtextbf{57}, 045003
  (2015).
  
  \bibitem[34]{Sanchis_2021}L.~Sanchis, M.~Garcia-Munoz, E.~Viezzer,
  A.~Loarte, L.~Li, Y.~Liu, A.~Snicker, L.~Chen, F.~Zonca, S.~Pinches, and
  D.~Zarzoso, {\newblock}Nuclear Fusion \tmtextbf{61}, 046006 (2021).
  
  \bibitem[35]{heidbrink2021}W.~W. Heidbrink and R.~B. White,
  {\newblock}Physics of Plasmas \tmtextbf{27}, 030901 (2020).
  
  \bibitem[36]{chen2019}L.~CHEN and F.~ZONCA, {\newblock}Plasma Science and
  Technology \tmtextbf{21}, 125101 (2019).
  
  \bibitem[37]{kramer2012}G.~J. Kramer, L.~Chen, R.~K. Fisher, W.~W.
  Heidbrink, R.~Nazikian, D.~C. Pace, and M.~A. Van~Zeeland, {\newblock}Phys.
  Rev. Lett. \tmtextbf{109}, 035003 (2012).
  
  \bibitem[38]{Todo2018}Y.~Todo, {\newblock}Rev. Mod. Plasma Phys.
  \tmtextbf{3}, 33 (2018).
  
  \bibitem[39]{janev1993}R.~K. Janev and J.~J. Smith, {\newblock}Atomic and
  Plasma-material Interaction Data for Fusion \tmtextbf{4} (1993).
  
  \bibitem[40]{todo2006}Y.~Todo, {\newblock}Phys. Plasmas (1994-present)
  \tmtextbf{13}, (2006).
  
  \bibitem[41]{boozer1981}A.~H. Boozer and G.~Kuo-Petravic, {\newblock}The
  Physics of Fluids \tmtextbf{24}, 851 (1981).
  
  \bibitem[42]{todo2014}Y.~Todo, M.~V. Zeeland, A.~Bierwage, and W.~Heidbrink,
  {\newblock}Nucl. Fusion \tmtextbf{54}, 104012 (2014).
\end{thebibliography}

\end{document}